\documentclass[prd,preprintnumbers,pacs,12pt]{revtex4}
\usepackage{epsfig}
\usepackage{graphicx}
\usepackage {axodraw}

\usepackage{bm}
\usepackage{amsmath}
\usepackage{amssymb}
\usepackage{amscd}
\usepackage{latexsym}

\newcommand{\be}{\begin{equation}}

\newcommand{\ee}{\end{equation}}
\newcommand{\bea}{\begin{eqnarray}}
\newcommand{\eea}{\end{eqnarray}}

\newcommand{\nn}{\nonumber}
\newcommand{\de}{\partial}
\def\Black{}
 \def\AliasBlue{}
 \def\Blue{}
 \def\Brown{}

\begin{document}

\newcommand{\bra}[1]{\langle #1|}
\newcommand{\ket}[1]{|#1\rangle}
\newcommand{\braket}[2]{\langle #1|#2\rangle}
\newcommand{\tr}{\textrm{Tr}}
\newcommand{\lag}{\mathcal{L}}
\newcommand{\mbf}[1]{\mathbf{#1}}
\newcommand{\desl}{\slashed{\partial}}
\newcommand{\Desl}{\slashed{D}}

\renewcommand{\bottomfraction}{0.7}
\newcommand{\epsi}{\varepsilon}

\newcommand{\nl}{\nonumber \\}
\newcommand{\tc}[1]{\textcolor{#1}}
\newcommand{\sla}{\not \!}
\newcommand{\spinor}[1]{\left< #1 \right>}
\newcommand{\cspinor}[1]{\left< #1 \right>^*}
\newcommand{\Log}[1]{\log \left( #1\right) }
\newcommand{\Logq}[1]{\log^2 \left( #1\right) }
\newcommand{\mr}[1]{\mathrm{#1}}
\newcommand{\cw}{c_\mathrm{w}}
\newcommand{\sw}{s_\mathrm{w}}
\renewcommand{\i}{\mathrm{i}}
\renewcommand{\Re}{\mathrm{Re}}
\newcommand{\yText}[3]{\rText(#1,#2)[][l]{#3}}
\newcommand{\xText}[3]{\put(#1,#2){#3}}

\title{Drell-Yan production at the LHC in a four site Higgsless model}

\author{Elena Accomando$^{(a)}$, Stefania De Curtis$^{(a)}$, Daniele Dominici$^{(a,b)}$ and Luca Fedeli$^{(b)}$}
\affiliation{ $(a)$ INFN, 50019 Sesto F., Firenze, Italy,
 $(b)$ Department of Physics, University of Florence, 50019 Sesto F., Firenze, Italy.}


\begin{abstract}
We consider a four site Higgsless model based on the $SU(2)_L\times
SU(2)_1\times SU(2)_2\times U(1)_Y$ gauge symmetry, which predicts two neutral
and four charged extra gauge bosons, $Z_{1,2}$ and $W^\pm_{1,2}$. We compute the
properties of the new particles, and derive indirect and direct limits on
their masses and couplings from LEP and Tevatron data. In contrast to other
Higgsless models, characterized by fermiophobic extra gauge bosons, here
sizeable fermion-boson couplings are allowed by the electroweak precision
data. The prospects of detecting the new predicted particles in the favoured
Drell-Yan channel at the LHC are thus investigated. The outcome is that all
six extra gauge bosons could be discovered in the early stage of the LHC
low-luminosity run.

\end{abstract}
\pacs{12.60.Cn, 11.25.Mj, 12.39.Fe}

\maketitle

\def\to{\rightarrow}
\def\ptl{\partial}
\def\beq{\begin{equation}}
\def\eeq{\end{equation}}
\def\bea{\begin{eqnarray}}
\def\eea{\end{eqnarray}}
\def\nn{\nonumber}
\def\half{{1\over 2}}
\def\rhalf{{1\over \sqrt 2}}
\def\calo{{\cal O}}
\def\call{{\cal L}}
\def\calm{{\cal M}}
\def\del{\delta}
\def\eps{\epsilon}
\def\lam{\lambda}

\def\anti{\overline}
\def\delfac{\sqrt{{2(\del-1)\over 3(\del+2)}}}
\def\heff{h'}
\def\square{\boxxit{0.4pt}{\fillboxx{7pt}{7pt}}\hspace*{1pt}}
    \def\boxxit#1#2{\vbox{\hrule height #1 \hbox {\vrule width #1
             \vbox{#2}\vrule width #1 }\hrule height #1 } }
    \def\fillboxx#1#2{\hbox to #1{\vbox to #2{\vfil}\hfil}   }

\def\braket#1#2{\langle #1| #2\rangle}
\def\gev{~{\rm GeV}}
\def\gam{\gamma}
\def\sn{s_{\vec n}}
\def\sm{s_{\vec m}}
\def\mm{m_{\vec m}}
\def\mn{m_{\vec n}}
\def\mh{m_h}
\def\sumn{\sum_{\vec n>0}}
\def\summ{\sum_{\vec m>0}}
\def\vl{\vec l}
\def\vk{\vec k}
\def\ml{m_{\vl}}
\def\mk{m_{\vk}}
\def\gp{g'}
\def\gt{\tilde g}
\def\hw{{\hat W}}
\def\hz{{\hat Z}}
\def\ha{{\hat A}}

\def\yy{{\cal Y}_\mu}
\def\yyt{{\tilde{\cal Y}}_\mu}
\def\lq{\left [}
\def\rq{\right ]}
\def\dmu{\partial_\mu}
\def\dnu{\partial_\nu}
\def\dmus{\partial^\mu}
\def\dnus{\partial^\nu}
\def\gp{g'}
\def\gpt{{\tilde g'}}
\def\gs{g''}
\def\ggs{\frac{g}{\gs}}
\def\eps{{\epsilon}}
\def\tr{{\rm {tr}}}
\def\V{{\bf{V}}}
\def\W{{\bf{W}}}
\def\Wt{\tilde{ {W}}}
\def\Y{{\bf{Y}}}
\def\Yt{\tilde{ {Y}}}
\def\L{{\cal L}}
\def\s{s_\theta}
\def\st{s_{\tilde\theta}}
\def\c{c_\theta}
\def\ct{c_{\tilde\theta}}
\def\gt{\tilde g}
\def\et{\tilde e}
\def\At{\tilde A}
\def\Zt{\tilde Z}
\def\Wpt{{\tilde W}^+}
\def\Wmt{{\tilde W}^-}

\newcommand{\Apt}{{\tilde A}_1^+}
\newcommand{\Bpt}{{\tilde A}_2^+}
\newcommand{\Amt}{{\tilde A}_1^-}
\newcommand{\Bmt}{{\tilde A}_2^-}
\newcommand{\Wtp}{{\tilde W}^+}
\newcommand{\Atp}{{\tilde A}_1^+}
\newcommand{\Btp}{{\tilde A}_2^+}
\newcommand{\Atm}{{\tilde A}_1^-}
\newcommand{\Btm}{{\tilde A}_2^-}
\def\mathswitchr#1{\relax\ifmmode{\mathrm{#1}}\else$\mathrm{#1}$\fi}
\newcommand{\Pe}{\mathswitchr e}
\newcommand{\Pp}{\mathswitchr {p}}
\newcommand{\PZ}{\mathswitchr Z}
\newcommand{\PW}{\mathswitchr W}
\newcommand{\PD}{\mathswitchr D}
\newcommand{\PU}{\mathswitchr U}
\newcommand{\PQ}{\mathswitchr Q}
\newcommand{\Pd}{\mathswitchr d}
\newcommand{\Pu}{\mathswitchr u}
\newcommand{\Ps}{\mathswitchr s}
\newcommand{\Pc}{\mathswitchr c}
\newcommand{\Pt}{\mathswitchr t}
\newcommand{\rd}{{\mathrm{d}}}
\newcommand{\GW}{\Gamma_{\PW}}
\newcommand{\GZ}{\Gamma_{\PZ}}
\newcommand{\GeV}{\unskip\,\mathrm{GeV}}
\newcommand{\MeV}{\unskip\,\mathrm{MeV}}
\newcommand{\TeV}{\unskip\,\mathrm{TeV}}
\newcommand{\fba}{\unskip\,\mathrm{fb}}
\newcommand{\pba}{\unskip\,\mathrm{pb}}
\newcommand{\nba}{\unskip\,\mathrm{nb}}
\newcommand{\PT}{P_{\mathrm{T}}}
\newcommand{\PTmiss}{P_{\mathrm{T}}^{\mathrm{miss}}}
\newcommand{\CM}{\mathrm{CM}}
\newcommand{\inv}{\mathrm{inv}}
\newcommand{\sig}{\mathrm{sig}}
\newcommand{\tot}{\mathrm{tot}}
\newcommand{\evt}{\mathrm{evt}}
\def\mathswitch#1{\relax\ifmmode#1\else$#1$\fi}
\newcommand{\M}{\mathswitch {M}}
\newcommand{\MW}{\mathswitch {M_\PW}}
\newcommand{\MZ}{\mathswitch {M_\PZ}}
\newcommand{\Mt}{\mathswitch {M_\Pt}}
\def\si{\sigma}
\def\beqar{\begin{eqnarray}}
\def\eeqar{\end{eqnarray}}
\def\refeq#1{\mbox{(\ref{#1})}}
\def\reffi#1{\mbox{Fig.~\ref{#1}}}
\def\reffis#1{\mbox{Figs.~\ref{#1}}}
\def\refta#1{\mbox{Table~\ref{#1}}}
\def\reftas#1{\mbox{Tables~\ref{#1}}}
\def\refse#1{\mbox{Sect.~\ref{#1}}}
\def\refses#1{\mbox{Sects.~\ref{#1}}}
\def\refapps#1{\mbox{Apps.~\ref{#1}}}
\def\refapp#1{\mbox{App.~\ref{#1}}}
\def\citere#1{\mbox{Ref.~\cite{#1}}}
\def\citeres#1{\mbox{Refs.~\cite{#1}}}

\def\Black{}
 \def\AliasBlue{}
 \def\Blue{}
 \def\Brown{}

\section{Introduction}
During the last years a remarkable activity has been devoted to investigate
Higgsless  models
\cite{Csaki:2003dt,Agashe:2003zs,Csaki:2003zu,Barbieri:2003pr,Nomura:2003du,
Cacciapaglia:2004zv,Cacciapaglia:2004rb,Cacciapaglia:2004jz,Contino:2006nn}
because they emerge in a natural way when considering local gauge theories in
five dimensions (5D). Their major outcome consists in delaying the unitarity
violation of vector boson scattering (VBS) amplitudes to higher energies
compared to the answer of the Standard Model (SM) without a light Higgs, via
the exchange of Kaluza-Klein (KK) excitations \cite{SekharChivukula:2001hz}.
The discretization of the compact fifth dimension to a lattice generates the
so-called deconstructed theories which are chiral lagrangians with a number of
replicas of the gauge group equal to the number of lattice sites
\cite{ArkaniHamed:2001ca,Arkani-Hamed:2001nc,Hill:2000mu,Cheng:2001vd,Abe:2002rj,Falkowski:2002cm,Randall:2002qr,Son:2003et,deBlas:2006fz}.
Models have been proposed, assuming a $SU(2)_L\times SU(2)_R\times U(1)_{B-L}$
gauge group in the 5D bulk,
\cite{Csaki:2003dt,Agashe:2003zs,Csaki:2003zu,Barbieri:2003pr,
Cacciapaglia:2004zv,Cacciapaglia:2004rb,Cacciapaglia:2004jz,Contino:2006nn},
in the framework suggested by the AdS/CFT correspondence, or also with a
simpler gauge group $SU(2)$ in the bulk
\cite{Foadi:2003xa,Hirn:2004ze,Casalbuoni:2004id,Chivukula:2004pk,Georgi:2004iy}.

The drawback of all these models, as with technicolor theories, is to reconcile
the presence of a relatively low KK-spectrum, necessary to delay the unitarity
violation to TeV-energies, with the electroweak precision tests (EWPT) whose
measurements can be expressed in terms of the $\eps_1,\eps_2$ and $\eps_3$
(or $T, U, S$) parameters. More in detail, while $\eps_1$ and $\eps_2$ are
protected by the custodial symmetry
\cite{Peskin:1990zt,Peskin:1992sw,Altarelli:1991zd,Altarelli:1998et}, shared
by both the aforementioned classes of models, the $\eps_3$ ($S$) parameter
constitutes the real obstacle to EWPT consistency. This problem can be solved
by either delocalizing fermions along the fifth dimension
\cite{Cacciapaglia:2004rb,Foadi:2004ps} or, equivalently in the deconstructed
version of the model, by allowing for direct couplings between new vector
bosons and SM fermions \cite{Casalbuoni:2005rs}. In the simplest version of
this latter class of models, corresponding to just three lattice sites and
gauge symmetry $SU(2)_L\times SU(2)\times U(1)_Y$ (the so-called BESS model
\cite{Casalbuoni:1985kq,Casalbuoni:1986vq}), the requirement of vanishing of
the $\eps_3$ parameter implies that the new triplet of vector bosons is almost
fermiophobic. As a consequence, the only production channels where the new
gauge bosons can be searched for are those driven by boson-boson couplings.
The Higgsless literature has been thus mostly focused on difficult
multi-particle processes which require high luminosity to be detected, that is
vector boson fusion (VBF) and associated production of new gauge bosons with
SM ones \cite{Birkedal:2004au,Belyaev:2007ss,He:2007ge}.

We extend the minimal three site model by inserting an additional lattice
site. This new four site Higgsless model, based on the
$SU(2)_L\times SU(2)_1\times SU(2)_2\times U(1)_Y$ gauge symmetry, predicts
two neutral and four charged extra gauge bosons, $Z_{1,2}$ and $W^\pm_{1,2}$,
and satisfies the EWPT constraints without necessarily having fermiophobic
resonances. Within this framework, the more promising Drell-Yan processes
become particularly relevant for the extra gauge boson search at the LHC.

In Section \ref{linear} we review the main properties of the model, in
particular we derive the couplings of new gauge bosons to SM fermions. In
Section \ref{bounds}, we discuss the bounds on masses and couplings coming
from EWPT and partial wave unitarity requirement. In
Section \ref{dy}, we discuss the prospects of detection of the new particles
in Drell-Yan channels at the LHC, then we give out our conclusions. In
Appendix \ref{appendixA}, we
compute the spectrum of the new gauge bosons and in Appendix \ref{numbers}
we list numerical values corresponding to the considered
scenarios.

\section{Review of the model}
\label{linear}

The class of models we are interested in follows the idea of
dimensional deconstruction
\cite{ArkaniHamed:2001ca,Arkani-Hamed:2001nc,Hill:2000mu,Cheng:2001vd}
and  was recently studied in \cite{Casalbuoni:2005rs}. The
so-classified theories can also be seen as generalizations of the
BESS model
\cite{Casalbuoni:1985kq,Casalbuoni:1986vq,Casalbuoni:1989xm} to an
arbitrary number of new triplets of gauge bosons. In their general
formulation
\cite{Foadi:2003xa,Hirn:2004ze,Casalbuoni:2004id,Chivukula:2004pk,Georgi:2004iy},
they are based on the  $SU(2)_L\otimes SU(2)^K\otimes U(1)_Y$ gauge
symmetry, and contain $K+1$ non linear $\sigma$-model scalar fields
$\Sigma_i$, ${i=1,\cdots ,K+1}$, transforming as \bea
&&\Sigma_1\to L\Sigma_1 U_1^\dagger,\nn\\
&&\Sigma_i\to U_{i-1}\Sigma_i U_i^\dagger\
,\,\,\,\,\,\,\,i=2,\cdots,K,
\nn\\
&&\Sigma_{K+1}\to U_K\Sigma_{K+1} R^\dagger,
\label{transfs}
\eea
with $U_i\in
SU(2)$, $L\in SU(2)_L$, $R\in U(1)_Y$. The
related Lagrangian  for scalars and gauge fields
is given by \be {\cal L}=\sum_{i=1}^{K+1}f_i^2{\rm
Tr}[D_\mu\Sigma_i^\dagger D^\mu\Sigma_i]-\frac 1 2\sum_{i=1}^K{\rm
Tr}[(F_{\mu\nu}^i)^2] -\frac 1 2{\rm Tr}[(F_{\mu\nu}(\Wt))^2] -\frac
1 2{\rm Tr}[(F_{\mu\nu}(\Yt))^2] , \label{lagrangian:l}
\ee
with the covariant derivatives  defined as follows
\bea
&D_\mu\Sigma_1=\de_\mu\Sigma_1-i\gt \Wt_\mu\Sigma_1+i\Sigma_1 g_1
\tilde{A}_\mu^1,&\nn\\
&D_\mu\Sigma_i=\de_\mu\Sigma_i-ig_{i-1}\tilde{A}_\mu^{i-1}\Sigma_i+i\Sigma_i
g_i \tilde{A}_\mu^i,&\,\,\,\,\,\,\,i=2,\cdots,K\nn\\
&D_\mu\Sigma_{K+1}=\de_\mu\Sigma_{K+1}-ig_{K}\tilde{A}_\mu^{K}\Sigma_{K+1}+i
\gpt\Sigma_{K+1}\Yt_\mu,& \label{covderivative} \eea where
$\tilde{A}_\mu^i=\tilde{A}_\mu^{ia}\tau^a/2$ and $g_i$ are the gauge
fields and gauge coupling constants associated to the groups $G_i$,
$i=1,\cdots ,K$;  $\Wt_\mu=\Wt_\mu^{a}\tau^a/2$,
$\Yt_\mu=\yyt\tau^3/2$ and $\gt$, $\gpt$ are the gauge fields and
couplings  associated to $SU(2)_L$ and $U(1)_Y$ respectively, and
$f_i$ are $K+1$ free parameters, the link coupling constants. In the continuum
limit, different $f_i$ can describe a generic warped metric, while
the flat case can have a correspondence with a scenario where all $f_i$ are
equal.

Direct couplings of new gauge bosons to SM fermions can be included in a way
that preserve the symmetry of the model.
The fermion Lagrangian is given by \cite{Casalbuoni:2005rs}
\bea
{\cal L}_{fermions}&=&\bar\psi_L i\gamma^\mu \de_\mu \psi_L +
\bar\psi_R i\gamma^\mu \de_\mu \psi_R\nn\\
&-&\frac 1 {{1+\sum^K_{i=1} b_i}}\bar\psi_L \gamma^\mu
 \gt\Wt_\mu\psi_L\nn\\
&-&
\sum_{i=1}^K  \frac {b_i}{{1+\sum^K_{j=1} b_j}}\bar\psi_L \gamma^\mu
 g_i \tilde{A}_\mu^i\psi_L\nn\\
&-&\bar\psi_R\gamma^\mu  (\gpt \Yt_\mu+ \frac 1 2 \gpt(B-L)
\yyt)\psi_R-\bar\psi_L \gamma^\mu\frac 1 2 \gpt (B-L)
\yyt\psi_L\label{eq:13} \eea where $b_i$ are arbitrary dimensionless
parameters, which we assume universal, and $\psi_{L(R)}$ denote the standard quarks and leptons.
Direct couplings of the new gauge bosons
to the right-handed fermions can also be introduced
 but they are strongly constrained by data from non-leptonic K-decays
and $b\to s\gamma$ processes \cite{Yao:2006px} to be of order 10 $^{-3}$ \cite{Bechi:2006sj}.

We will not include these couplings here. Fermion mass terms can be
built using the field $U=\Sigma_1 \Sigma_2\cdots \Sigma_{K+1}$
transforming as $U\to LUR^\dagger$ \cite{Casalbuoni:2005rs}. The
case $K=1$ corresponds to the BESS model
\cite{Casalbuoni:1985kq,Casalbuoni:1986vq} whose phenomenology was
studied at the  LHC  in Ref. \cite{Casalbuoni:1990uf}. This model
has been recently rediscovered as a three site Higgsless model
\cite{SekharChivukula:2006cg}, and its phenomenology has now
received a renewed attention
\cite{Contino:2006nn,He:2007ge,Belyaev:2007ss,Birkedal:2004au}. The
requirement of consistency of the model with electroweak precision
data can be satisfied only via a strong cancelation between
gauge-mixing and $b$-parameter contributions to $\epsilon_3$ ($S$),
ending up with fermiophobic vector resonances.

In this paper we concentrate on the case $K=2$ where the EWPT constraints can
be satisfied without having necessarily fermiophobic resonances. The Drell-Yan
channels become relevant here. For simplicity we will assume $g_1=g_2$ and
$f_1=f_3$. This choice corresponds to a L-R symmetry in the new
gauge sector, leading to a definite parity for the corresponding
gauge bosons once the standard gauge interactions are turned off.

Summing up, the parameters of the model are $g_1$, $f_1$, $f_2$,
$b_1$ and $b_2$. For the gauge sector, this model corresponds to the one
proposed in Ref. \cite{Casalbuoni:1989xm} once fixed the set of parameters
to be: $a=0$, $b=c=d/2=2f_1^2/v^2$,
$g^{\prime\prime}=\sqrt{2}g_1$, $v=(\sqrt{2}G_F)^{-1/2}$.
In Appendix \ref{appendixA}, we give the charged and neutral gauge bosons
spectrum in the four site model.

\subsection{Fermion gauge boson couplings}
Using the explicit relations between gauge fields ($\tilde W, \tilde A^{1,2}$)
and mass eigenvectors ($W, W^\pm_{1,2}$) given in Eqs.
(\ref{W}-\ref{A2c}), we get the charged fermion-boson interaction:
\begin{equation}\label{CC}
{\mathcal L}_{CC}=\bar{\psi_L} \gamma^\mu T^- \psi_L \left(a_W
W_{\mu}^+ +a_{1}^c W_{1\mu }^+ +a_{2}^c W_{2\mu }^+\right) + h.c.
\end{equation}
with
\be
a_W=-\frac{\gt}{\sqrt{2}}\left(1-\frac{b}{2}\right)\left(1-\frac{\gt^2}{g_1^2}\frac{z_W}{2}\right)
\ee
\be\label{a1c}
a_1^c=-\frac{g_1}{2(1+b_+)}\left(b_+-\frac{\gt^2}{g_1^2}\right)
\ee
\be\label{a2c}
a_2^c=-\frac{g_1}{2(1+b_+)}\left(b_- -\frac{\gt^2}{g_1^2}z^2\right)
\ee
where
\be\label{b}
z_W=\frac 1 2 (1+z^4) \quad\quad b=\frac{b_+-b_-z^2}{(1 +b_+) }
\quad\quad b_\pm =b_1\pm b_2 \quad\quad
z=\frac{f_1}{\sqrt{f_1^2+2f_2^2}} \ee and we have neglected terms
${\mathcal O}(\gt^4/g_1^4)$ and ${\mathcal O}(b_i\gt^2/g_1^2)$. The
reason why we neglect terms ${\mathcal O}(b_i\gt^2/g_1^2)$ is that
the direct fermion couplings to the new resonances are of the
order $b_i g_1$ and we do not expect couplings bigger than the SM
ones to be allowed, that is  $b_i\leq \gt/g_1$. This is confirmed by
the bounds from LEP discussed hereafter. Notice also that $0\leq
z<1$ and, from Eq. (\ref{A7}), $M_1=z M_2<M_2$.

In analogous way, using the relations between gauge fields ($\tilde
 {W}^3, \tilde{Y},\tilde A^3_{1,2}$) and neutral eigenvectors ($A, Z,
Z_{1,2}$) given in Eqs. (\ref{W30}-\ref{A230}) we derive the neutral
fermion-boson interactions:
\begin{equation}\label{NC}
{\mathcal L}_{NC}=\bar{\psi} \gamma^\mu (-a_F \mathbf{Q} A_\mu +a_1^n
Z_{1\mu}+ a_2^n Z_{2\mu} + a_Z Z_\mu )\psi
\end{equation}
with \be\label{aF2} a_F=\gt s_
{\tilde{\theta}}\left(1-\frac{\gt^2}{g_1^2}z_\gamma\right)\equiv e
\ee \be a_Z=-\frac{\gt}{c_
{\tilde{\theta}}}\left(1-\frac{b}{2}\right)\left(1-\frac{\gt^2}{g_1^2}\frac{z_Z}{2}\right)\left[\mathbf{T^3}-\frac{s^2_
{\tilde{\theta}}}{\left(1-\frac{b}{2}\right)}\left(1-\frac{\gt^2c_
{\tilde{\theta}}}{g_1^2s_
{\tilde{\theta}}}z_{Z\gamma}\right)\mathbf{Q}\right] \ee
\be\label{a1n2} a_1^n=-\frac{g_1}{\sqrt{2}(1+b_+)}\left(b_+
-\frac{\gt^2}{g_1^2} \frac{c_{ 2\tilde{\theta}}}{c^2
_{\tilde{\theta}}}\right)\mathbf{T^3}+\frac{\gt^2 \tan^2
\tilde{\theta}}{\sqrt{2} g_1} \mathbf{Q} \ee \be\label{a2n2}
a_2^n=-\frac{g_1}{\sqrt{2}(1+b_+)}\left(b_-
-\frac{\gt^2}{g_1^2}\frac{z^2}{c^2 _{\tilde{\theta}}}
\right)\mathbf{T^3}-\frac{ \gt^2 z^2
\tan^2\tilde{\theta}}{\sqrt{2}g_1 }\mathbf{Q} \ee where
\be\label{zZF} z_\gamma=s^2_ {\tilde{\theta}},\quad\quad
z_Z=\frac{1}{2}\frac{(z^4+c^2_{2\tilde{\theta}})}{c^2_{\tilde{\theta}}},
\quad\quad
z_{Z\gamma}=-\tan \tilde{\theta}c_{ 2\tilde{\theta}} \ee and
$\mathbf{T^3}=\tau^3_L/2$ ($\tau^3_L\psi_L=\pm \psi_L$ and
$\tau^3_L\psi_R=0$), $\tan \tilde{\theta}=
s_{\tilde{\theta}}/c_{\tilde{\theta}}={\gpt}/{\gt}$, $\mathbf{Q}$ is
the electric charge in unit $e$ (the proton charge).

\section{Bounds from EWPT and from perturbative Unitarity}
\label{bounds} Simplest Higgsless models suffer for a tension
between EWPT and perturbative unitarity. This tension can be
alleviated by allowing for  delocalization of fermions in the fifth
dimension \cite{Cacciapaglia:2004rb,Foadi:2004ps} or by allowing for
direct couplings of the new gauge bosons to standard matter
\cite{Casalbuoni:2005rs}. We will explore this second possibility in
the four site model.

\subsection{Bounds from EWPT}

In order to get bounds on the parameter space of the model, it is
convenient to derive the new physics contribution to the electroweak
parameters $\epsilon_1$, $\epsilon_2$ and $\epsilon_3$ (or $S$, $T$
and $U$) which are strongly constrained by the electroweak
measurements \cite{Barbieri:2004qk}.

These parameters can be
obtained from
$\Delta r_W$, $\Delta\rho$ and $\Delta k$ \cite{Altarelli:1993sz,Altarelli:1998et}:
\bea
\eps_1&=& \Delta\rho\,\nn\\
\eps_2 &=& \c^2\Delta\rho+\frac{\s^2}{c_{2\theta}}\Delta r_W-2 \s^2\Delta k\,\nn\\
\eps_3 &=& \c^2\Delta\rho+c_{2\theta}\Delta k \,
\label{epsdef}
\eea
with $\Delta r_W$ defined by:
\be
\frac{M^2_W}{M^2_Z}=
 \c^2\left[1-\frac{\s^2}{c_{2\theta}}\Delta r_W\right]\,
\ee
$\Delta\rho$ and $\Delta k$
 given in terms of the neutral current
couplings to the $Z$ gauge boson
\be
\L^{neutral}(Z)=-\frac{e}{\s\c}
\Big(1+\frac{\Delta\rho}{2}\Big)Z_\mu\overline\psi [\gamma^\mu
g_V+\gamma^\mu \gamma_5g_A]\psi
\ee
with
\be
g_V =
\frac{\mathbf{T^3}}{2}-s^2_{\theta_{eff}} \mathbf{Q},\quad g_A =
-\frac{\mathbf{T^3}}{2},\quad s^2_{\theta_{eff}} = (1+\Delta k) \s^2\,
\ee
and $\s $ defined by: \be\label{Sin2} \s^2\c^2=\frac{{\sqrt{2}}
e^2}{8 M_Z^2G_F}.
\ee
Therefore, in this scheme, the physical inputs
are chosen to be the electric charge, the Fermi constant and the $Z$-boson
mass.

As already said, the parameters of the model are $g_1$, $f_1$, $f_2$, $b_1$ and
$b_2$. Fixing $M_Z$, given in our model by Eq. (\ref{A14}), to its
experimental value, we get a relation among the five initial parameters which
allows to express $g_1$ in terms of the others. This implies that the model
has four independent free parameters. By Eq.
(\ref{A7}), we prefer to use the masses instead of the link couplings; we
thus end up with $M_1$, $M_2$, $b_1$ and $b_2$ as free parameters. The
explicit relation for $g_1$ is (at leading order):
 \be\label{g1}
g_1\sim\frac{e}{s_{ 2\theta}}\frac{M_1}{M_Z}\sqrt{2 (1-z^2)}.
\ee
Let
us compute the expression for the Fermi constant $G_F$ in our model.
Neglecting terms ${\mathcal O}(\gt^4/g_1^4)$ and
${\mathcal O}(b_i\gt^2/g_1^2)$. We get
\be\label{GF} \frac{G_F}{\sqrt{2}}=\frac{a_W^2}{4
M_W^2}+\frac{(a_1^c)^2}{4 M_{1,c}^2}+\frac{(a_2^c)^2}{4 M_{2,c}^2}=
\frac{\gt^2}{8M_W^2}
\left(1-\frac{b}{2}\right)^2\left(1-\frac{\gt^2}{g_1^2}z_W\right)+\frac{\beta}{4}
\ee with \be\label{f22}
\beta=\frac{(1-z^2)(b_+^2+b_-^2z^2)}{8(1+b_+)^2f^2}, \quad\quad
f=\frac {f_1f_2}{\sqrt{f_1^2+2 f_2^2}} \ee and $M_W$, $M_{1,c}$,
$M_{2,c}$, given in Eqs. (\ref{Mw}-\ref{M2c}). Using also Eq. (\ref{A14}) we get the relation between $M_W$  and $M_Z$
\be M_W^2=M_Z^2 c^2_{\tilde{\theta}}\left(1-\frac{e^2}{g_1^2
s^2_{\tilde{\theta}}}(z_W-z_Z)\right) \label{mwmz} \ee where we have
used Eq. (\ref{aF2}), again neglecting terms ${\mathcal
O}(\gt^4/g_1^4)$ and ${\mathcal O}(b_i\gt^2/g_1^2)$. Requiring the
tree-level  SM relation in Eq. (\ref{Sin2}) to hold true also in our
model, we derive the relation between the $\tilde{\theta}$ angle and
the physical inputs: \be
s^2_{\tilde{\theta}}=\frac{1}{2}\left(1-\sqrt{1-s^2_{2\theta}
X}\right) \label{sint} \ee where \be
X=\frac{1}{R}\left[1+\frac{4e^2}{g_1^2}\left(\frac{Rz_W}{s^2_{2\theta}}-1\right)\right],\quad\quad
R=\frac{1-\frac{\beta }{2\sqrt{2}G_F}}{\left(1-\frac{b}{2}\right)^2}
\ee By using the definitions of $\Delta\rho$, $\Delta r_W$, $\Delta
k$ and Eqs. (\ref{GF}, \ref{mwmz}, \ref{sint}) we get, again neglecting terms
${\mathcal O}(\gt^4/g_1^4)$ and ${\mathcal O}(\gt^2/g_1^2 b_i)$
\bea\label{rhoK}
\Delta \rho&=&2\left(-1+\sqrt{1-\frac{\beta }{2\sqrt{2}G_F}}\right)\nn \\
\Delta k&=&-1
+\frac{s^2_{\tilde{\theta}}}{s^2_{\theta}}\frac{1}{\left(1-\frac{b}{2}\right)}\left[1-\frac{e^2}{g_1^2
s^2_{\tilde{\theta}}} z_{Z\gamma}\right] \eea

\be \Delta r_W=\frac{c_
{2\theta}}{s^2_{\theta}}\left[1-\frac{c^2_{\tilde{\theta}}}{c^2_{\theta}}\left(1-\frac{e^2}{g_1^2
s^2_{\tilde{\theta}}}(z_W-z_Z)\right)\right] \ee We can now compute
the new physics contribution to the $\epsilon_{1,2,3}$ by using Eq.
(\ref{epsdef}) as in Ref. \cite{Casalbuoni:2005rs}. We find: \be
\epsilon_{1,2}=-\frac{(1-z^2)(b_+^2+z^2b_-^2)}{4},
\quad\quad\epsilon_3=\frac{g^2}{2g_1^2}(1-z^4)-\frac{b}{2} \ee with
$b$ and $b_\pm$ given in Eq. (\ref{b}) and $g=e/s_\theta$. As
previously mentioned, while $\eps_3$ is strongly affected by the
direct fermion-boson couplings, having a linear dependence on the
$b$-parameter, $\eps_1$ and $\eps_2$ receive a mild contribution.
They display in fact only a quadratic behaviour in $b_\pm$, owing to
the $SU(2)$ custodial symmetry.

\begin{figure}[!htb]
\begin{center}
\includegraphics[width=6 cm]{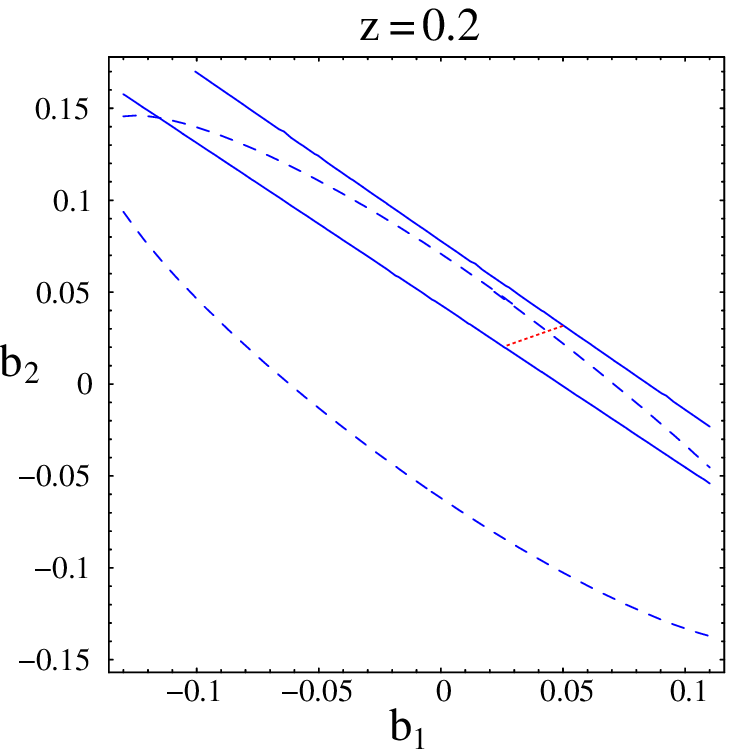}~~~~~~~~~~~~~
\includegraphics[width=6 cm]{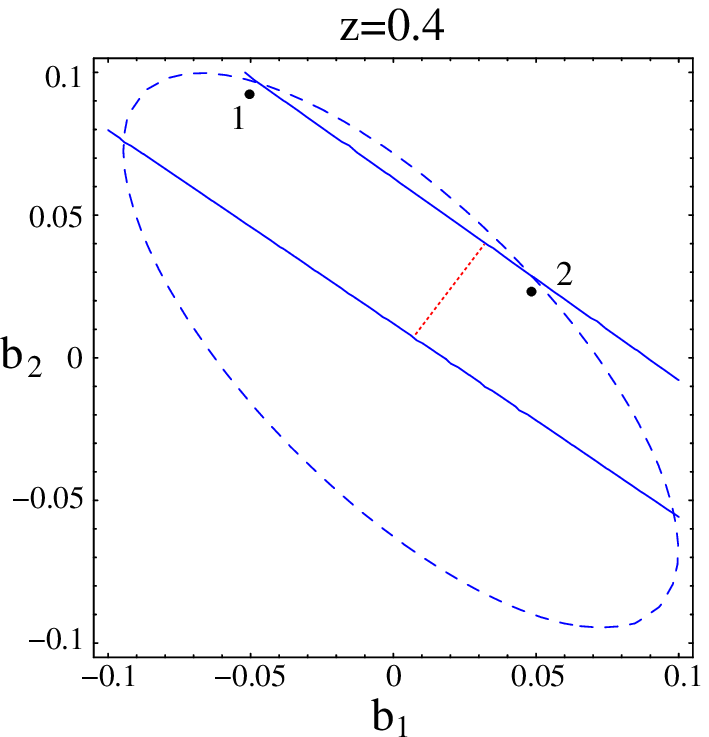}
\end{center}
\caption{95\% C.L. bounds on the plane $(b_1,b_2)$ from $\eps_1$
(dash line) and $\eps_3$ (solid line). Left panel: $z=0.2$ and
$450\le M_1(GeV)\le550$.
Right panel: $z=0.4$,
$500\le M_1(GeV)\le 1000$.
The allowed regions are the
internal ones.
The red dots represent the so-called ideal cancellation. Also shown
are the points corresponding to the scenarios 1 and 2 of the
following phenomenological analysis.
}\label{bi1}
\end{figure}

\begin{figure}[!htb]
\begin{center}
\includegraphics[width=6 cm]{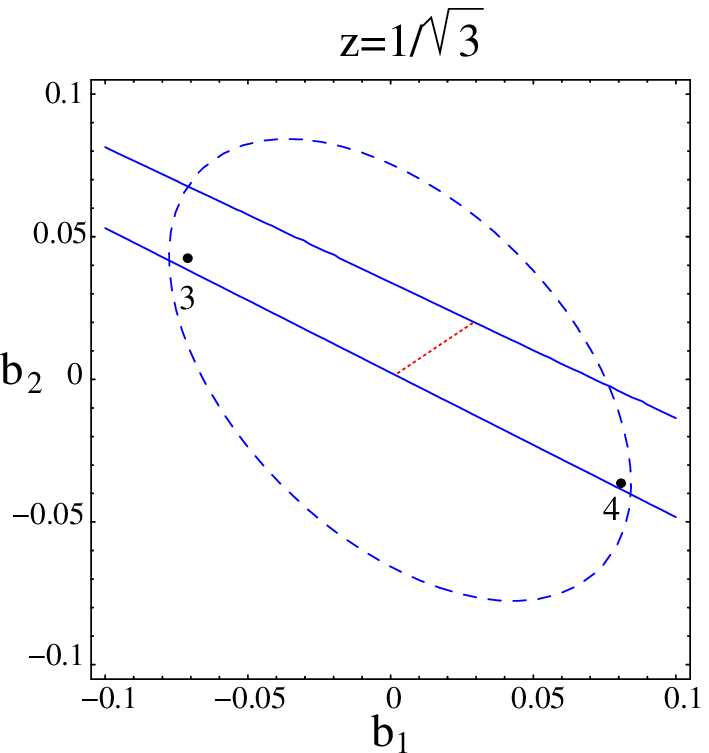}~~~~~~~~~~~~~
\includegraphics[width=6 cm]{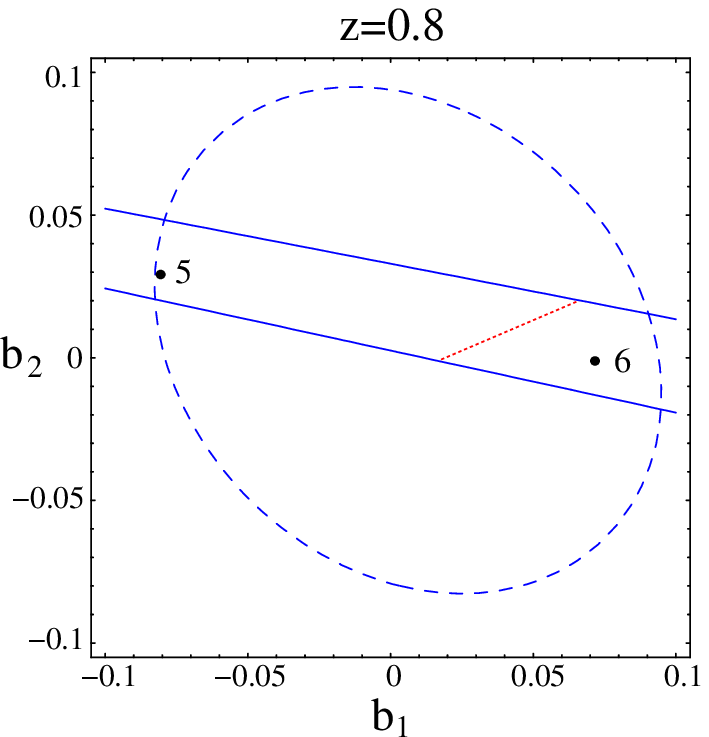}
\end{center}
\caption{Same as Fig \ref{bi1}. Left panel:  $z=1/\sqrt{3}$ (corresponding to
$f_1=f_2$),
$700\le M_1(GeV)\le 1750$.
Right panel:  $z=0.8$,
$700\le M_1(GeV)\le 1600$. Also shown are the points corresponding
to the scenarios 3, 4, 5 and 6 of the following phenomenological
analysis.}
\label{bi2}
\end{figure}

In order to get bounds on the model parameters we have to compare
with the experimental values for the $\eps_{1,2,3}$
\cite{Barbieri:2004qk}: \be \epsilon_1^{exp}=(5.0 \pm 1.1)\times
10^{-3}, \quad \epsilon_2^{exp}=(-8.8 \pm 1.2)\times 10^{-3},\quad
\epsilon_3^{exp}=(4.8 \pm 1.0)\times 10^{-3} \ee after adding to the
new physics contribution the radiative corrections. In our analysis,
we have only considered the SM radiative corrections evaluated for
$m_t=172.5$ GeV and an heavy Higgs boson, $m_H=1$ TeV, taken as a
cut-off: \be \epsilon_1^{rad}=3.4 \times 10^{-3},\quad
\epsilon_2^{rad}=-6.5 \times 10^{-3},\quad \epsilon_3^{rad}= 6.7
\times 10^{-3}.\quad \ee In Figs.~\ref{bi1} and \ref{bi2}, we show
the 95$\%$ C.L. experimental bounds from $\epsilon_{1}$ and
$\epsilon_3$ ($\epsilon_2$ doesn't give any relevant limit thanks to
its negative experimental value) in the plane $(b_1,b_2)$ for
different choices
 of heavy gauge boson masses, $M_{1,2}$, at fixed
ratio $z=M_1/M_2$ (see Eq. \ref{A7}). The bounds are obtained by the
standard $\chi$-square fit \be
\chi_i^2=\left({\eps_i+\eps_i^{rad}-\eps_i^{exp}\over{\sigma_i^{exp}}}\right
)^2=5.99 ~~~~ with ~ i=1,3 \ee which has been calculated exactly in
the $b_i$ parameters. As an example of the dependence of the
$\eps_i$ constraints on the free parameter $z$, we consider four
values $z=(0.2, 0.4, 1/\sqrt{3}, 0.8)$ and show the corresponding
results in the four plots of Figs.~\ref{bi1} and \ref{bi2}. From
there, it is clear that for $z=0.2$ there is a little match between
experimental data and model parameters, instead for $z\geq 0.4$
there is a non negligible allowed
strip. For each choice of $z$ we choose $M_{1,2}$ to vary inside the
region allowed by the perturbative unitarity limit discussed in the
second part of this  section. The bounds from $\epsilon_1$ are quite
insensitive to the value of the resonance masses, those from
$\epsilon_3$ goes from up to down by increasing $M_1$. Also, the range of low
 $M_{1}$ values for given $z$ has been cut in order to be consistent with
our neglecting of terms of $O(\gt^4/g_1^4)$ and taking into account Eq.~(\ref{g1}).

It is instructive to translate these bounds on the plane delimited
by the direct couplings between the two charged (neutral) gauge
bosons, $W^\pm_{1,2} (Z_{1,2})$, and ordinary matter. In fact, the
following expression for $\epsilon_3$ can be easily derived by using
Eqs. (\ref{a1c}), (\ref{a2c}), (\ref{a1n2}), (\ref{a2n2}) and
neglecting terms ${\mathcal O}(\gt^4/g_1^4)$ and ${\mathcal
O}(b_i\gt^2/g_1^2)$:
\be \epsilon_3\sim\frac{1}{g_1}(a_1^c-z^2
a_2^c)\sim-\frac{\sqrt{2}}{g_1} ({a_{1L}^e-z^2
a_{2L}^e})-\frac{e^2}{g_1^2 c^2_\theta} (1+z^4)
\label{epsilon3} \ee
where $a_{1,2}^c$ are the fermion couplings of the charged
$W_{1,2}^{\pm}$-boson and $a_{iL}^{e} (i=1,2)$ are the couplings of
the neutral $Z_{1,2}$-boson to the left-handed electron component.
>From the previous relations we note that the contribution of the
fermion couplings of $Z_2$ and $W_2^{\pm}$ are multiplied by a
factor $z^2$. As a consequence, in order to satisfy the stringent
bounds from $\eps_3$, fermion couplings of the $Z_1$ and $W_1^\pm$
resonances will be much more
constrained with respect to the $Z_2$ and $W_2^\pm$ ones.\\

\begin{figure}[]
\begin{center}
\unitlength1.0cm
 \begin{picture}(8,7)
   \put(-3.3,0.6){\epsfig{file=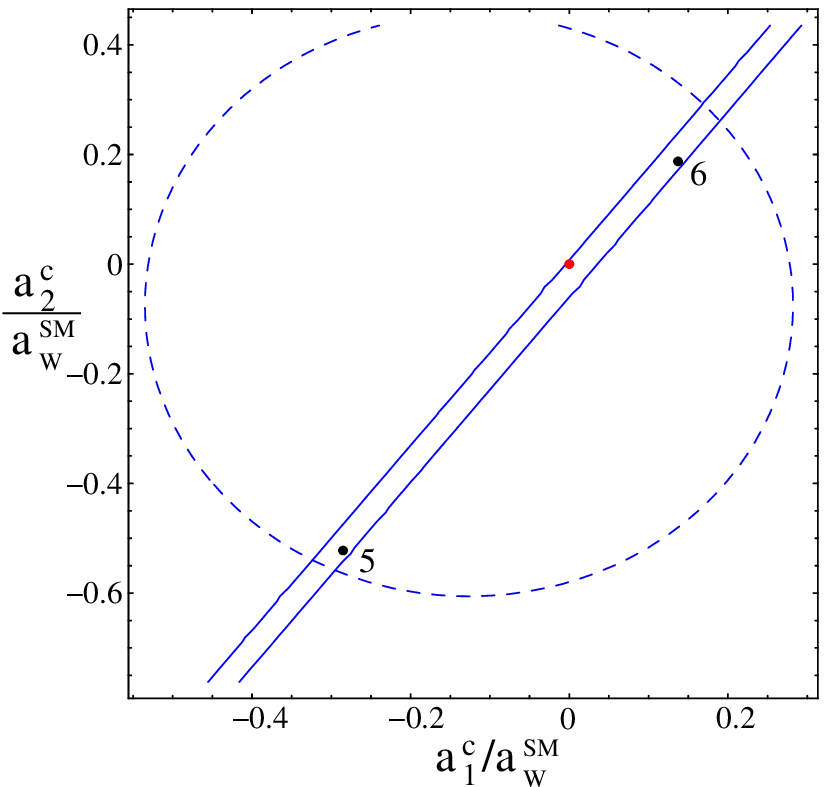,width=6.6cm}}
   \put(4.3,0.4){\epsfig{file=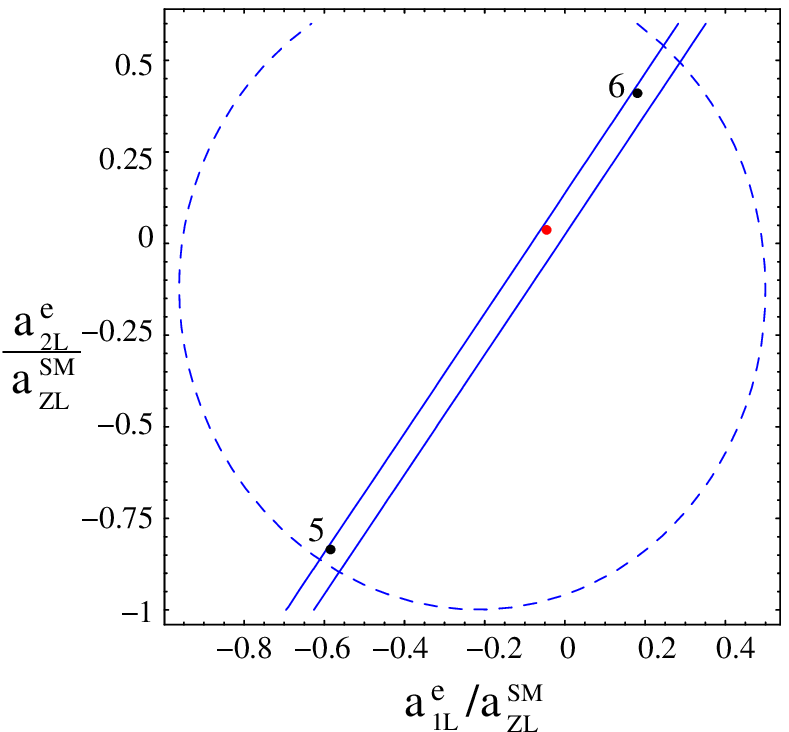,width=7.cm}}
 \end{picture}
\end{center}
\caption{{ 95\% C.L. bounds from   $\epsilon_1$ (dashed line) and
$\epsilon_3$ (solid line), for $M_1=1$ TeV and $z=0.8$ on the plane
$(a_{1}^c/a^{SM}_W,a_2^c/a^{SM}_W)$ with $a_W^{SM}=-e/(\sqrt{2}
s_\theta)$ (left panel), on the plane
$(a_{1L}^e/a_{ZL}^{SM},a_{2L}^e/a_{ZL}^{SM})$ with
$a_{ZL}^{SM}=-e/(s_\theta c_\theta)(1/2-s^2_\theta)$ (right panel).
In both cases the central dot represents the so-called ideal
cancellation.  Also shown
are the points corresponding to the scenarios 5 and 6 of the
following phenomenological analysis.}}\label{fig1}
\end{figure}

Let us derive the experimental bounds from $\eps_1$ and $\eps_3$ for
the charged and neutral fermion couplings normalized with respect to
the SM ones. The results are shown in Fig.~\ref{fig1} for $M_1=1$
TeV and $z=0.8$. The measurement of $\eps_3$ strongly constraints
the physical space of the couplings, instead the measurement of
$\epsilon_1$ gives looser bounds. The central dot represents the
point where $a_1^c=a_2^c=0$, the so-called ideal cancellation
\cite{Casalbuoni:2005rs,Chivukula:2005ji,SekharChivukula:2005cc,Bechi:2006sj,Casalbuoni:2007xn}.
This is the assumption under which the most recent phenomenological
analysis have been performed at the LHC within the three site
Higgsless model \cite{Belyaev:2007ss, He:2007ge}. There, in fact,
the new resonances are forced by EWPT to be fermiophobic (or almost
fermiophobic). In the four site extension instead, while the
relation between the couplings of the two gauge-boson triplets with
SM-fermions is strongly constrained by the $\epsilon_3$-parameter,
their magnitude is weakly limited by $\epsilon_1$. As a result, the
direct fermion-boson couplings can be of the same order of the SM
ones, as shown in Fig.~\ref{fig1}. The phenomenological consequence
is that, while the minimal Higgsless model can be explored only in
very complex multi-particle processes (triple gauge boson production
or vector boson fusion channels) which require high luminosity, the
four site Higgsless model could be proved in the more promising
Drell-Yan channel already at the LHC start-up.

We are not considering in this paper new physics loop corrections to
the $\epsilon$ parameters: it has been recently shown that in some
Higgsless model this effect can be important for reconciling these
schemes  with electroweak precision data
\cite{Dawson:2007yk,Barbieri:2008cc,Abe:2008hb}.

We have instead considered the limits from LEP2 by computing the $\hat S, \hat T, \hat U, W, Y, V$ and X \cite{Barbieri:2004qk}. The result is that $\hat T, \hat U$, and $V$ vanish due to the SU(2) custodial symmetry, $X, Y$ and $W$ are of order $b_i\tilde g^2/g_1^2$ or $\tilde g^4/g_1^4$ and therefore do not alter the stringent limits already obtained from $\epsilon_3$.

\subsection{Perturbative unitarity bounds}

As well known, in the SM without a light Higgs the vector boson scattering
(VBS) amplitudes violate
perturbative unitarity at an energy scale of the order of
$\Lambda_{\rm HSM}\simeq 1.7\TeV$ \cite{Lee:1977eg}.
One of the strong motivations for Higgsless models, derived from 5D local
gauge theories, is their ability to delay such a unitarity
violation to higher energies, via the exchange of the predicted extra
gauge bosons.

In this section, we briefly discuss the issue of partial wave unitarity. The
corresponding bound can be easily obtained by using the Equivalence Theorem
relating, at high energy, the gauge boson scattering amplitudes to the
corresponding Goldstone ones
\cite{Lee:1977eg,Cornwall:1973tb,Vayonakis:1976vz,Chanowitz:1985hj}.
Let us put $\gt=\gt'=0$ and $K=2$ in the  Lagrangian given in Eq.
(\ref{lagrangian:l}) and expand the fields as
$\Sigma_i=\exp[{i\alpha_i\vec{\pi_i}\cdot\vec{\tau}}]$. We get:
\bea\label{3.93} \sum_{i=1}^{3}
f_i^2\mathbf{Tr}\left[D_\mu\Sigma_i^\dag D^\mu\Sigma_i\right]&\sim &
\sum_{i=1}^{3}\Big[ 2f_i^2\alpha_i^2
\partial_\mu\vec{\pi}_i\cdot\partial^\mu\vec{\pi}_i
+\frac{2}{3}\alpha_i^4f_i^2\left[(\vec{\pi}_i\cdot\partial_\mu\vec{\pi}_i)^2
-\vec{\pi}_i^2(\partial_\mu\vec{\pi}_i)^2\right]\Big]\nonumber\\
&& +\sum_{i=1}^2g_{i}\vec{\tilde A}_{i}^\mu\cdot
\left(f_i^2\alpha_i\partial_\mu\vec{\pi}_i-f_{i+1}^2\alpha_{i+1}\partial_\mu\vec{\pi}_{i+1}\right)+\cdots
\eea
The unitary gauge for the $\tilde A_i$ gauge bosons is given by
\be
\alpha_i=\frac{f}{2f_i^2},  \quad {\vec\pi}_i(x)={\vec\pi}(x)
\ee with
$f$ given in Eq. (\ref{f22}), which ensures  a canonical kinetic term
for the $\pi$'s:
\be
\sum_{i=1}^32f_i^2\alpha_i^2\partial_\mu\vec{\pi}_i\cdot
\partial^\mu\vec{\pi}_i=\frac{1}{2}\partial_\mu\vec{\pi}\cdot\partial^\mu\vec{\pi}.
\ee
We can now easily compute the scattering amplitude ${\mathcal
A}_{W_L^+W_L^-\rightarrow W_L^+W_L^-}$ which, for $\sqrt{s}>> M_W$, is equal
to ${\mathcal A}_{\pi^+\pi^-\rightarrow\pi^+\pi^-}$ due to the Equivalence
Theorem. This amplitude for high energy ($\sqrt{s}>> M_{1,2}$) is dominated by
the four-pion vertex which can be extracted from Eq. (\ref{3.93}).
Comparing the four-linear Lagrangian coefficient in Eq. (\ref{3.93}), computed
in unitary gauge, with the analogous interaction given in the SM without a
light Higgs, we get
\be
{1\over{6v_{SM}^2}}={f^4\over{24}}\sum_{i=1}^3{1\over{f_i^6}}.
\ee
The pion-pion scattering amplitudes can thus be written in our model as:
\be {\mathcal
A}_{\pi^+\pi^-\rightarrow\pi^+\pi^-}\sim
-\frac{f^4}{4}\sum_{i=1}^3\frac{u}{f_i^6}
\ee
where $u$ is the Mandelstam variable, $u=-s(1-\cos\theta )/2$, with $s$
the CM-energy squared and $\theta$ the $\pi^-$ scattering angle. We also get:
\be
{\mathcal A}_{\pi^3\pi^3\rightarrow\pi^3\pi^3}\sim 0,~~~~~~{\mathcal
A}_{\pi^+\pi^-\rightarrow\pi^3\pi^3}\sim
\frac{f^4}{4}\sum_{i=1}^3\frac{s}{f_i^6}
\ee
which are related to the physical scattering amplitudes
$Z_LZ_L\rightarrow Z_LZ_L$ and $W^+_LW^-_L\rightarrow Z_LZ_L$, respectively.

\begin{figure}[t] \centerline{
\epsfxsize=7cm\epsfbox{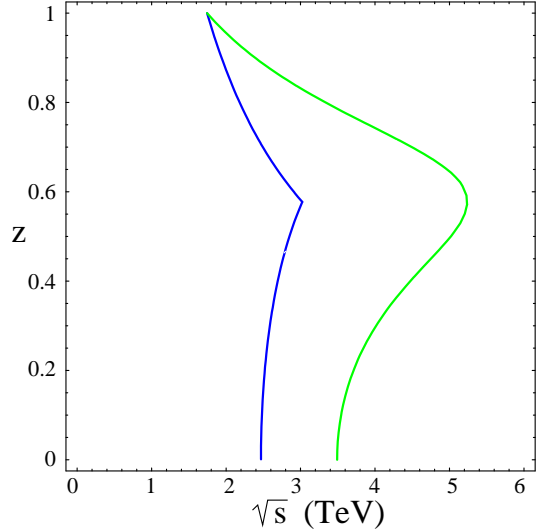} } \caption {{Partial wave
unitarity bounds on the plane ($\sqrt{s}$, $z$), by requiring
$|a_0|\leq 1$ for scattering amplitudes with external SM gauge bosons
(green lighter curve), and for scattering amplitudes with both SM and extra
gauge bosons as external states (blue darker curve). The allowed regions are
on the left of the curves.
\label{a0} }}
\end{figure}

By considering the zero-isospin partial wave matrix for all VBS amplitudes
with SM longitudinal gauge bosons as external states
\be
a_0={1\over{16\pi}}{s\over{v_{SM}^2}}\left(\begin{array}{cc}
{1\over 2} & {1\over{\sqrt{2}}} \\
{1\over{\sqrt{2}}} & 0
\end{array}\right)
\ee
with normalization factors for the two channels $W^+_LW^-_L$ and
$(1/\sqrt{2})Z_LZ_L$ as in Ref. \cite{Lee:1977eg},
and requiring the partial wave bound $\vert a_0\vert= 1$ for the maximum
eigenvalue, we get the result shown by the
green (lighter) curve in Fig.~\ref{a0}. The maximum energy scale, up to which
the perturbative unitarity can be delayed, is reached for $z=1/\sqrt{3}$ or
equivalently $f_1=f_2$ which can be interpreted as a flat-metric scenario. In
this particular case, such
a delay with respect to the answer of the SM without a light Higgs is
modulated by the factor $(K+1)\equiv 3$:
$\Lambda_{\rm{four-site}}=3~\Lambda_{\rm HSM}\sim 5$ TeV.
However, the four site model has in addition two vector-boson triplets with,
potentially, bad behaving longitudinal scattering amplitudes. In order to be
predictive, one has to require a fully perturbative regime for all involved
particles. The unitarity limit must thus be extended, in order to ensure a
good high energy behaviour for all scattering amplitudes, i.e. with both SM
and extra gauge bosons as external states.

For energies much higher than the masses of the new vector bosons
($\sqrt{s}>> M_{1,2}$), we can determine such a unitarity bound by
considering the eigenchannel amplitudes corresponding to all
possible longitudinal vector bosons. This can be done again by using
the Equivalence Theorem and evaluating the chiral Lagrangian in the
unitary gauge for all the vector bosons given by
$\Sigma_i=\exp[{i\vec{\pi_i}\cdot\vec{\tau}/(2 f_i)]}$. In this
gauge, the amplitudes are diagonal and the high energy result is
simply given by: \be {\mathcal
A}_{\pi_i^+\pi_i^-\rightarrow\pi_i^+\pi_i^-}\sim -\frac{u}{4f_i^2},
~~~{\mathcal A}_{\pi_i^3\pi_i^3\rightarrow\pi_i^3\pi_i^3}\sim
0,~~~{\mathcal A}_{\pi_i^+\pi_i^-\rightarrow\pi_i^3\pi_i^3}\sim
\frac{s}{4f_i^2}. \ee Requiring $\vert a_0\vert=1$ for the maximum
eigenvalue, we get the blue (darker) curve shown in Fig.~\ref{a0}.
We see that, also in this case, the most delayed unitarity limit is
realized for the choice $f_1=f_2$ that is $z=1/\sqrt{3}$, and
corresponds to the rescaling $\Lambda^{\rm
total}_{\rm{four-site}}=\sqrt{K+1}~\Lambda_{\rm HSM}\sim 3~\TeV$.
The four site Higgsless model can thus preserve perturbative
unitarity over practically the whole effective energy range of the
LHC. The spectrum of the new predicted particles $W^\pm_{1,2}$ and
$Z_{1,2}$ must lie within that region: $M_{1,2}\le\Lambda^{\rm
total}_{\rm{four-site}}$.

In our following phenomenological analysis, the most stringent unitarity bound
in Fig.~\ref{a0} is considered. We conclude by noting that the three site
unitarity limits are lower or equal to the ones of the four site model, given
by the blue (darker) line of Fig.~\ref{a0}, depending on the value of the
ratio $f_1/\sqrt{f_1^2+f_2^2}$.

\section{Drell-Yan production at the LHC}
\label{dy} We can now consider the production of the  six new gauge
bosons, $W^\pm_{1,2}$ and $Z_{1,2}$, predicted by the four site
Higgsless model at the LHC through Drell-Yan channels. Owing to the
introduction of direct couplings between ordinary matter and extra
gauge bosons, in addition to the usual indirect ones due to the
mixing, the experimental bounds from electroweak precision data on
the model parameters are indeed less stringent. As a consequence,
and in contrast with the existing fermiophobic Higgsless literature,
quite large couplings between SM fermions and extra gauge bosons are
allowed (see Fig.~\ref{fig1}).

\subsection{Processes and their computation}
\label{se:processes}

We analyze in detail two classes of processes,
\renewcommand{\labelenumi}{(\roman{enumi})}
\begin{enumerate}
\item $\Pp\Pp\to l^+l^-$, with $l=\Pe,\mu$,
\qquad and
\item $\Pp\Pp\to l\nu_l$, with $l=\Pe,\mu$.
\end{enumerate}
The first class is characterized by two isolated charged leptons in
the final state. The latter gives instead rise to one isolated
charged lepton plus missing energy. In our notation, $l\nu_l$
indicates both $l^-\bar\nu_l$ and $l^+\nu_l$. The aforementioned
neutral and charged Drell-Yan channels can involve the production of
two neutral extra gauge bosons, $Z_1$ and $Z_2$, and four charged
extra gauge bosons $W_1^\pm$ and $W_2^\pm$ as intermediate states,
respectively. Both classes of processes are described by the formula
\beqar \rd\si^{h_1 h_2}(P_1,P_2,p_f) = \sum_{i,j}\int\rd x_1 \rd
x_2~ f_{i,h_1}(x_1,Q^2)f_{j,h_2}(x_2,Q^2)
\,\rd\hat\si^{ij}(x_1P_1,x_2P_2,p_f), \eeqar where $p_f$ summarizes
the final-state momenta, $f_{i,h_1}$ and $f_{j,h_2}$ are the
distribution functions of the partons $i$ and $j$ in the incoming
hadrons $h_1$ and $h_2$ with momenta $P_1$ and $P_2$, respectively,
$Q$ is the factorization scale, and $\hat\si^{ij}$ represent the
cross sections for the partonic processes. Since the two incoming
hadrons are protons and we sum over final states with opposite
charges, we find
\beqar\label{eq:convol_neut}
\rd\si^{h_1h_2}(P_1,P_2,p_f) = \int\rd x_1 \rd x_2 &&\sum_{Q=u,c,d,s,b}
\Bigl[f_{\bar\PQ,\Pp}(x_1,Q^2)f_{\PQ,\Pp}(x_2,Q^2)\,\rd\hat\si^{\bar\PQ\PQ}
(x_1P_1,x_2P_2,p_f)
\nl&&{}
+f_{\PQ,\Pp}(x_1,Q^2)f_{\bar\PQ,\Pp}(x_2,Q^2)\,\rd\hat\si^{\PQ\bar\PQ}
(x_1P_1,x_2P_2,p_f)\Bigr]
\eeqar
and
\beqar\label{eq:convol_char}
\rd\si^{h_1h_2}(P_1,P_2,p_f) = \int\rd x_1 \rd x_2 &&\sum_{U=u,c}\sum_{D=d,s}
\Bigl[f_{\bar\PD,\Pp}(x_1,Q^2)f_{\PU,\Pp}(x_2,Q^2)\,\rd\hat\si^{\bar\PD\PU}
(x_1P_1,x_2P_2,p_f)
\nl&&{}
+f_{\bar\PU,\Pp}(x_1,Q^2)f_{\PD,\Pp}(x_2,Q^2)\,\rd\hat\si^{\bar\PU\PD}
(x_1P_1,x_2P_2,p_f)
\nl&&{}
+f_{\PD,\Pp}(x_1,Q^2)f_{\bar\PU,\Pp}(x_2,Q^2)\,\rd\hat\si^{\PD\bar\PU}
(x_1P_1,x_2P_2,p_f)
\nl&&{}
+f_{\PU,\Pp}(x_1,Q^2)f_{\bar\PD,\Pp}(x_2,Q^2)\,\rd\hat\si^{\PU\bar\PD}
(x_1P_1,x_2P_2,p_f)\Bigr]
\eeqar
for neutral and charged processes, respectively.

The tree-level amplitudes for the partonic processes have been generated by
means of {\tt PHACT} \cite{Ballestrero:1999md}, a set of routines based on the
helicity-amplitude formalism of \citere{Ballestrero:1994jn}. The matrix
elements have been inserted in the Monte Carlo event generator (MCEG)
{\tt FAST$\_$2f}, dedicated to Drell-Yan processes at the EW and QCD leading
order. {\tt FAST$\_$2f} can compute simultaneously the new-physics signal and
the SM background. It can generate cross-sections and distributions for
any observable, including any kind of kinematical cuts. The code is
moreover interfaced with {\tt PYTHIA} \cite{Sjostrand:2006za}. This feature
can allow more realistic analysis, once {\tt FAST$\_$2f} is matched
with detector simulation programs. This extension and relative study will be
performed soon.

\subsection{Numerical setup}
\label{se:setup}

For the numerical results presented here, we have used the following
input values \cite{Yao:2006px}: $M_Z=91.187\GeV$, $\GZ=2.512\GeV$, $\GW
=2.105\GeV$,  $\alpha(M_Z)=1/128.88$, $G_F = 1.166\times 10^{-5}$ GeV$^{-2}$.
Additional input parameters are the quark-mixing matrix elements
\cite{Hocker:2001xe}, whose values have been taken to be
$|V_{\Pu\Pd}|=|V_{\Pc\Ps}|=0.975$,
$|V_{\Pu\Ps}|=|V_{\Pc\Pd}|=0.222$, and zero for all other relevant
matrix elements. In our scheme, the weak mixing-angle and the
$W$-boson mass are derived quantities.
For the matrix element
evaluation, we adopt the fixed-width scheme. And, we use the CTEQ6L
\cite{Pumplin:2002vw} for the parton distribution functions at the
factorization scales:
\begin{equation}
Q^2=M_{\inv}^2(l^+l^-)
\end{equation}
and
\begin{equation}
Q^2={1\over 2}\left (\PT^2(l)+\PT^2(\nu_l )\right )
\end{equation}
for neutral and charged Drell-Yan processes, respectively, where $M_{\inv}$
denotes the invariant mass and $\PT$ is the transverse momentum. This scale
choice appears to be appropriate for the calculation of differential cross
sections, in particular for lepton distributions at high energy scales.

We have, moreover, implemented a general set of acceptance cuts,
appropriate for LHC analysis, defined as follows:
\begin{itemize}
\item {lepton transverse momentum $\PT(l)>20\GeV$},

\item {missing transverse momentum $\PTmiss> 20\GeV$ for
  $\Pp\Pp\to l\nu_l$},

\item {lepton pseudo-rapidity $|\eta_l |< 2.5$},
where $\eta_l=-\log\left (\tan\theta_l/2\right )$, and
  $\theta_l$ is the polar angle of the charged lepton $l$
with respect to the beam.
\end{itemize}
For the different processes considered, we have also used further
cuts which are described in due time. We present
results for the LHC at $\CM$ energy $\sqrt s=14\TeV$ and an integrated
luminosity from $L=100\pba^{-1}$ to $L=100\fba^{-1}$.

\subsection{Extra gauge boson production in Drell-Yan channels}
\label{se:boson_production}

In the following two subsections we analyze the production of
charged and neutral extra gauge bosons in Drell-Yan channels. We
consider three different choices of mass spectrum and two sets of
couplings to fermions for each choice inside the region allowed by
EWPT and unitarity bounds:

\begin{table}[h]
$z=0.4\hspace{4cm} z=1/\sqrt{3}\hspace{4cm}z=0.8$
\begin{center}
\begin{tabular}{|c|c|c|c|}
\hline & $M_{1,2} (\GeV)$&$b_{1,2}$&   $g_1$  \\
\hline \hline
1&500,1250 &-0.05,0.09&2.7\\
\hline 2&500,1250
&0.06,0.02& 2.7\\
\hline
\end{tabular}~~~~
\begin{tabular}{|c|c|c|c|}
\hline & $M_{1,2} (\GeV)$&$b_{1,2}$&   $g_1$  \\
\hline \hline
3& 1732,3000 &-0.07,0.04&8.1\\
\hline 4&1732,3000
&0.08,-0.04& 8.1\\
\hline \end{tabular}~~~~
\begin{tabular}{|c|c|c|c|}
\hline & $M_{1,2} (\GeV)$&$b_{1,2}$&   $g_1$  \\
\hline \hline
5&1000,1250 &-0.08,0.03&3.5\\
\hline 6&1000,1250
&0.07,0.0&3.5\\
\hline
\end{tabular}
\end{center}
\caption{Choices of parameters for the six considered scenarios. The
corresponding value of $g_1$ evaluated by Eq. (\ref{g1}) is also
given.}
\label{tab:scenarios}
\end{table}

These six examples give an idea of the possible scenarios predicted
by the four site Higgsless model. In the model in fact the ratio
between the gauge boson masses of the first and second triplet, i.e.
$z=M_1/M_2$, is a free parameter. Hence, the distance between the
two masses is arbitrary as well (actually the mass eigenvalues are
slightly different from $M_{1,2}$ due to corrections ${\cal
O}(\gt/g_1)^2$ as given in Appendix \ref{appendixA} and \ref{numbers}). We
have thus
chosen three cases, corresponding to $z=(0.4, 1/\sqrt{3}, 0.8)$, and
representing from left to right very distant resonances, the flat-metric
scenario, and a spectrum which tends to degeneracy by increasing $z$.

Also the magnitude of the couplings has a wide spectrum, as can be
seen for example in Fig.~\ref{fig1} for the case $z=0.8$ and $M_1=1$
TeV. In order to show the importance and the impact at the LHC of
the related direct fermion-boson couplings, we consider two points
in each plane $(b_1,\; b_2)$: the first in the region down on the
left of the allowed range, and the second high on right (see Figs.~
\ref{bi1},\ref{bi2}). These are reported in the six scenarios listed
above.

We are now ready to discuss numerical results for the  charged and
neutral  Drell-Yan channels.

\subsubsection{$Z_1$ and $Z_2$ production at the LHC}
\label{se:neutral_production}

In this section, we present some cross sections and distributions
for the leptonic process $\Pp\Pp\to l^+l^-$ with $l=\Pe,\mu$.  These
final states allow to analyze the production of the two neutral
extra gauge bosons, $Z_{1,2}$, predicted by the four site
Higgsless model.

General studies of Higgsless models have shown that the new strongly
interacting vector bosons are expected to be produced at high CM energies.
We thus select this kinematical configuration by imposing an additional
cut on the invariant mass of the lepton pair, i.e.
$M_{\inv}(l^+l^-)\ge 400\GeV$.

As an illustration of the behaviour and the impact of the new
predicted particles at the LHC, we have chosen to analyze the
following three differential cross-sections:
\renewcommand{\labelenumi}{(\roman{enumi})}
\begin{enumerate}
\item distribution in the invariant mass of the reconstructed
$Z_{1,2}$-boson, $M_{\inv}(l^+l^-)$,
\qquad
\item distribution of the scattering angle in the CM frame, $\cos\theta^*_l$,
 taken in the region under the peak of the new resonances, i.e.
 $|M_{\inv}(l^+l^-)-M_i|\le 3\Gamma_i$ ($i=1,2$),
\item forward-backward charge asymmetry versus the dilepton invariant mass.
\end{enumerate}

As mentioned in \refse{se:boson_production}, we show results for
three values of the $z$-parameter, $z=(0.4, 1/ \sqrt{3}, 0.8)$, and
various $b_{1,2}$-sets. The corresponding neutral fermion-boson
couplings are summarized in Table \ref{tab4} of Appendix \ref{numbers}.

\begin{figure}[t]
\begin{center}
\unitlength1.0cm
 \begin{picture}(8,7)
  \xText{-2.5}{5.3}{\small{(a)~~$M_{1,2}=(500, 1250)\GeV$}}
  \xText{-1.3}{-0.1}{$\mr{M}_{\inv}(l^+l^-) \normalsize [\GeV]$}
  \xText{-4.}{3.}{$N_{evt}$}
  \xText{5.2}{5.3}{\small{(b)~~$M_{1,2}=(1732, 3000)\GeV$}}
  \xText{6.7}{-0.1}{$\mr{M}_{\inv}(l^+l^-) \normalsize [\GeV]$}
  \xText{3.9}{3.}{$N_{evt}$}
  \xText{-2.5}{-0.85}{\small{(c)~~$M_{1,2}=(1000, 1250)\GeV$}}
  \xText{-1.3}{-6.3}{$\mr{M}_{\inv}(l^+l^-) \normalsize [\GeV]$}
  \xText{-4.}{-3.3}{$N_{evt}$}
  \xText{5.2}{-0.85}{\small{(d)~~$M_{1,2}=(1000, 1250)\GeV$}}
  \xText{6.7}{-6.3}{$\mr{M}_{\inv}(l^+l^-) \normalsize [\GeV]$}
  \xText{3.9}{-3.3}{$N_{evt}$}
  \put(-4.2,-0.7){\epsfig{file=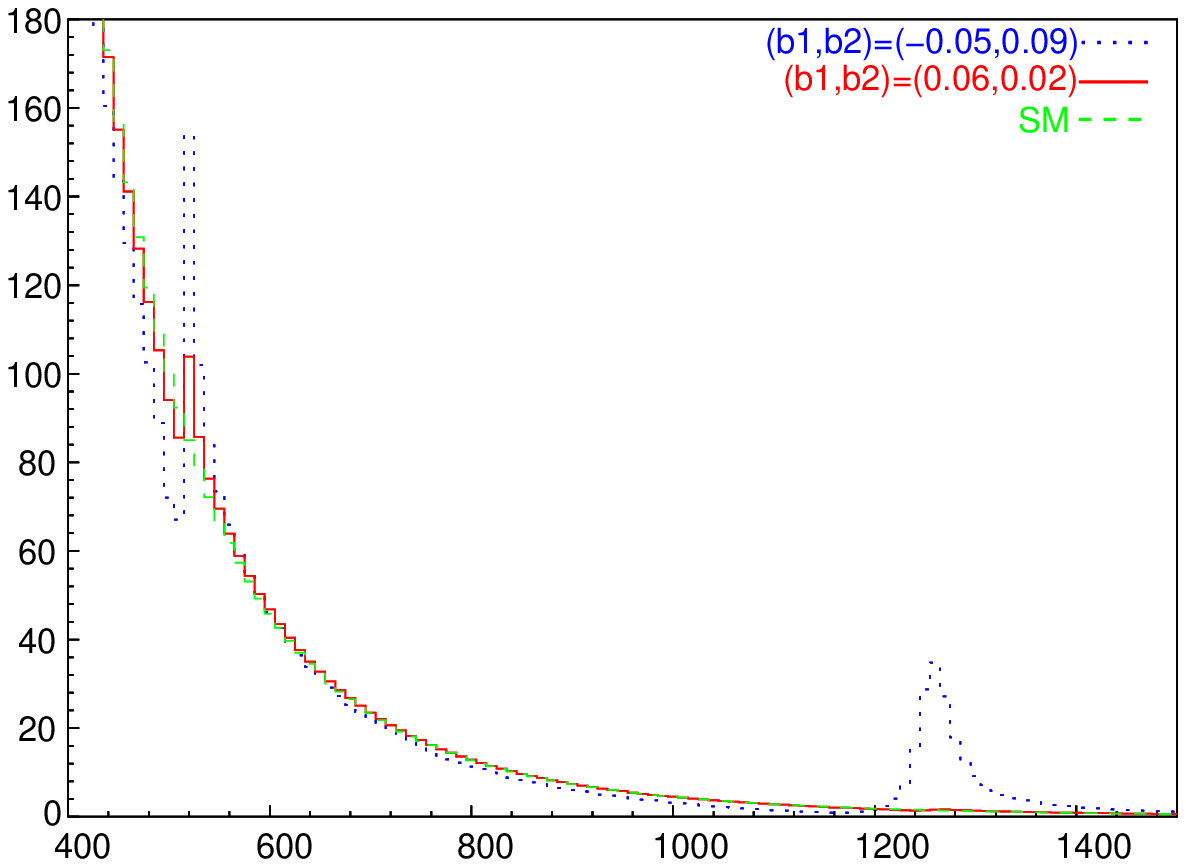,width=12.cm}}
  \put(3.6,-0.7){\epsfig{file=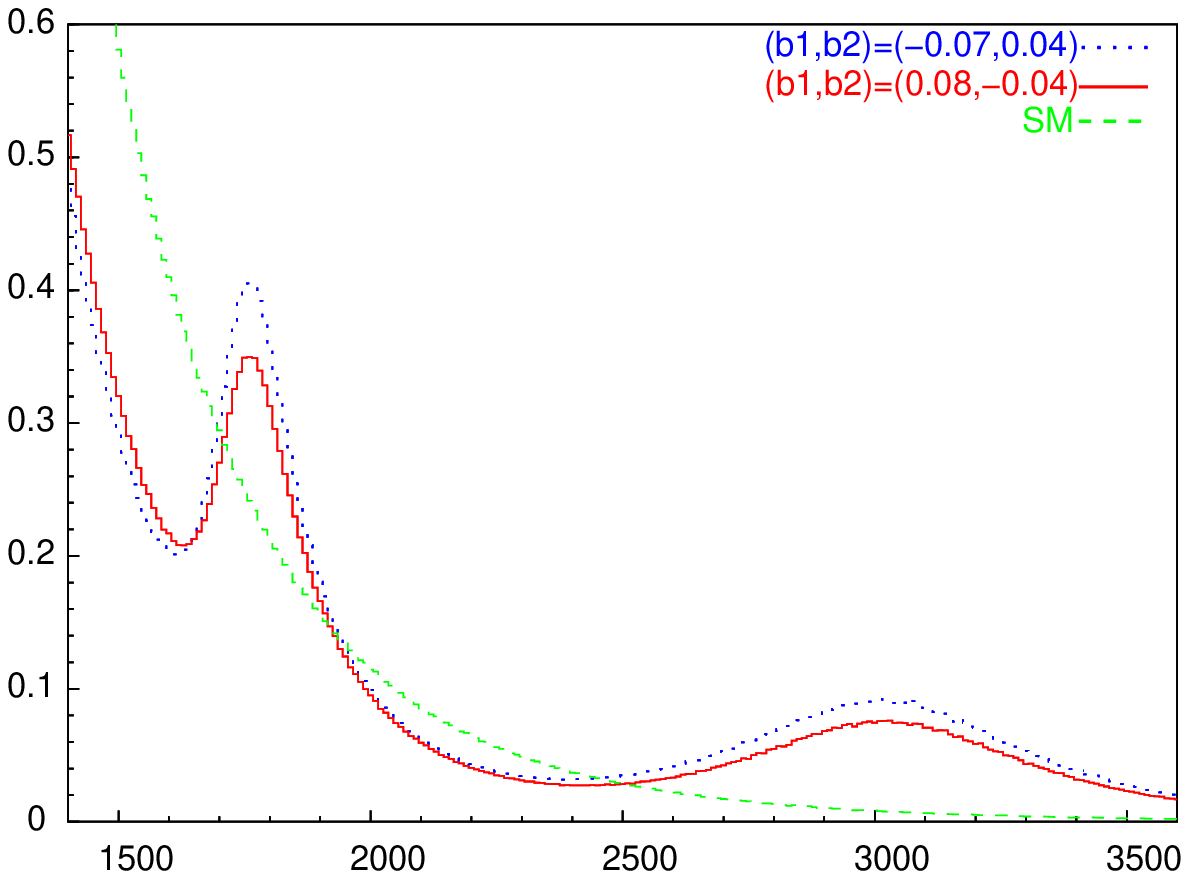,width=12.cm}}
  \put(-4.2,-6.9){\epsfig{file=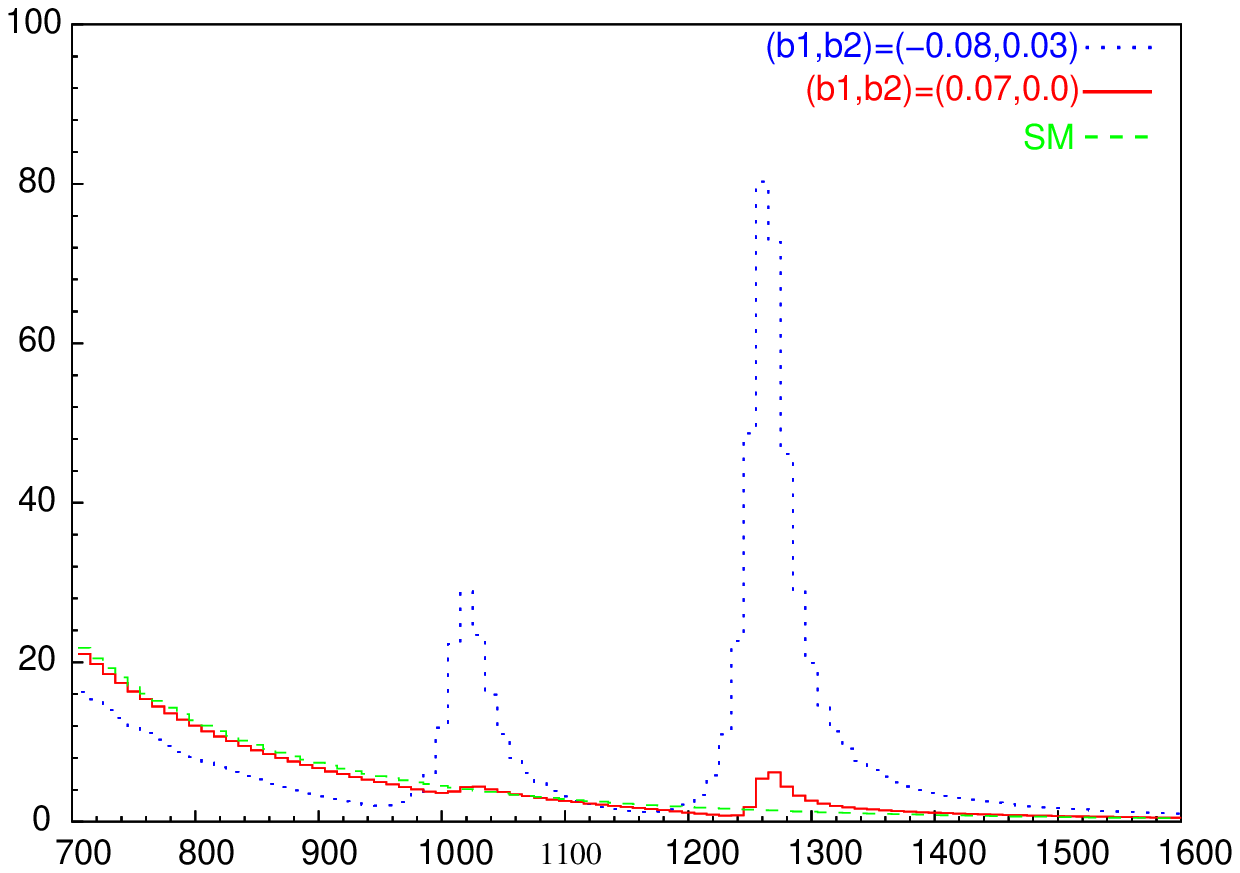,width=12.cm}}
  \put(3.6,-6.9){\epsfig{file=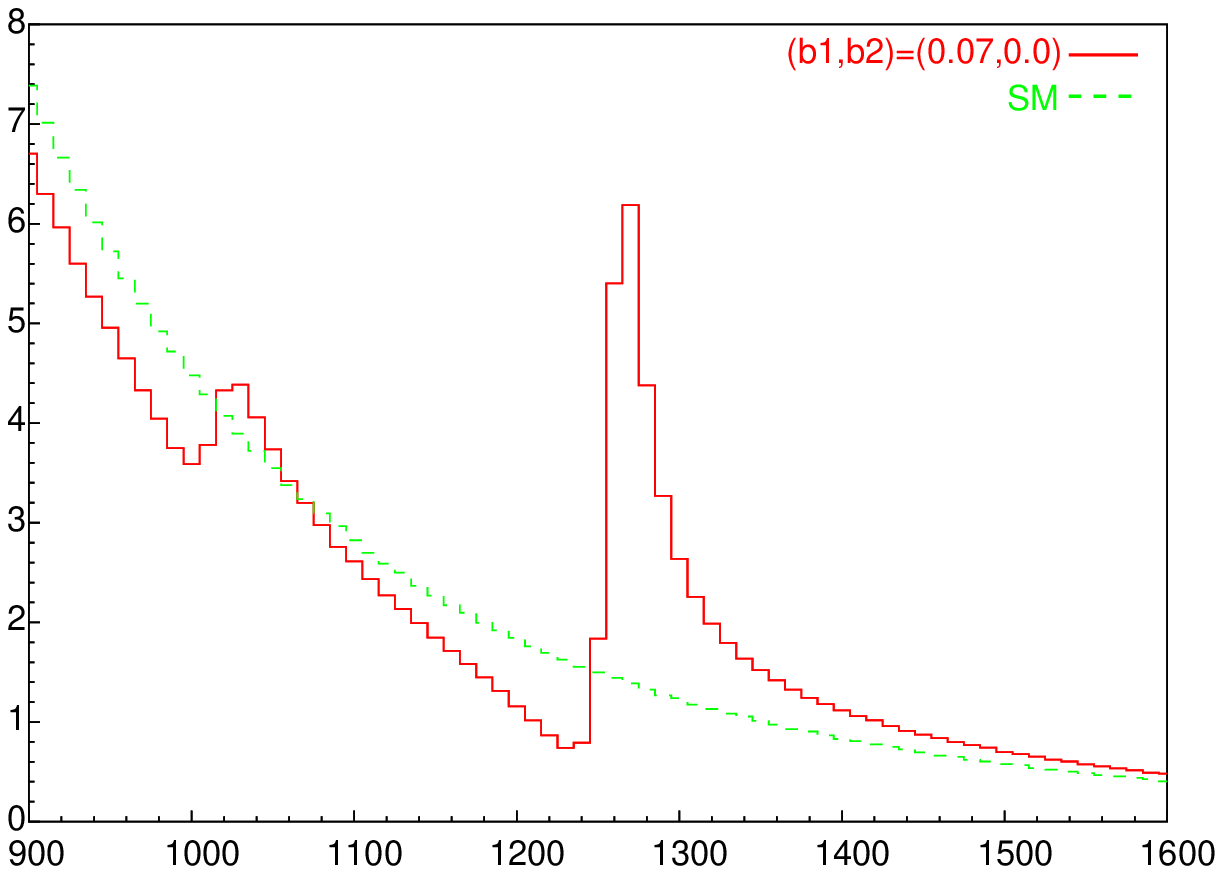,width=12.cm}}
 \end{picture}
\end{center}
\vskip 6.cm \caption{Total number of events in a 10$\GeV$-bin versus
the dilepton invariant mass, $M_{\inv}(l^+l^-)$, for the process
$\Pp\Pp\rightarrow l^+l^-$ at the integrated luminosity $L=10$
fb$^{-1}$ for the six scenarios of Table \ref{tab:scenarios}. We sum
over $e,\mu$. Standard cuts and legends as in the text.}
\label{fig:minv}
\end{figure}

We start from the spectrum of the neutral extra gauge bosons as it
could appear in the Drell-Yan channel at the LHC. In Fig. \ref{fig:minv}, we
plot the total number of events as a function of
the dilepton invariant mass, $M_{\inv}(l^+l^-)$, for the six
aforementioned scenarios. We have checked that all these cases are outside
the exclusion limit
from direct searches at the Tevatron with an integrated luminosity
L=4$\fba^{-1}$ \cite{tevatron-lumi}. From top to bottom, the three curves in
each plot represent the first $b_{1,2}$-setup, the latter, and the
SM prediction at fixed $M_{1,2}$ masses. We sum over $e,\mu$ and
apply standard acceptance cuts.

These examples show four peculiarities of the model. First of all,
the masses of the two neutral resonances $Z_{1,2}$ are not equally
spaced. They can be either very distant for low $z$-values (see
Figs.~\ref{fig:minv}a and \ref{fig:minv}b), or they can tend to be
almost degenerate for high $z$-values (see Fig.~\ref{fig:minv}c). For some
values of the $b_{1,2}$-parameter (see for example the case 6), one of
the two resonances could also disappear leaving a single-resonant
(or even not-resonant) spectrum like in Fig.~\ref{fig:minv}d. A
second feature is related to the interference between signal and
SM background. All four plots exhibit indeed a sizeable depletion of
the total number of events, compared to the SM prediction, in the
off-peak region. This characteristic is particularly evident by
increasing the new gauge boson masses (see Figs.~\ref{fig:minv}b and
\ref{fig:minv}c), and improves the detection rate. A further
distinctive behaviour is represented by the width magnitude. It
indeed increases with the fifth power of the extra gauge boson mass.
One can thus pass from configurations with very narrow resonances,
see Fig.~\ref{fig:minv}a, to scenarios characterized by broad peaks,
or even shoulders as in Fig.~\ref{fig:minv}b. The last feature
concerns the relative size of the $Z_{1,2}$ resonances. Tab.~\ref{tab4} in
Appendix \ref{numbers} shows that in most part of the
parameter space the $Z_1$-fermion couplings are smaller than the
$Z_2$-fermion ones. As a consequence, the height of the
$Z_1$-resonance is less pronounced than the $Z_2$-peak. This feature
can be washed out by the PDF effect, but it is clearly visible in
Fig.~\ref{fig:minv}c.

In order to have an idea of the detection rate expected at the LHC
for the Drell-Yan production of the extra $Z_{1,2}$ gauge bosons, in
Tab.~\ref{tab2} we have listed signal and total event number in the
two distinct on-peak regions
$|M_{\inv}(l^+l^-)-M_{1,2}|<\Gamma_{1,2}$ for the six considered
scenarios. From Tab.~\ref{tab2}, it is clear that while the second
$b_{1,2}$-setup, at each fixed $M_{1,2}$ mass, would need high
luminosity to be detected, the first $b_{1,2}$-setup could already
be visible at the LHC start-up with a luminosity L$\simeq$ 1 fb$^{-1}$.

\begin{table}
\begin{center}
\begin{tabular}{|c|c|c|c|c|c|c|c|c|}
\hline &$M_{1,2} (\GeV)$ & $b_{1,2}$ & $N_{\evt}^{\sig}(Z_1)$ &
$N_{\evt}^{\tot}(Z_1)$ & $\sigma (Z_1)$ & $N_{\evt}^{\sig}(Z_2)$ &
$N_{\evt}^{\tot}(Z_2)$
& $\sigma (\PZ_2)$ \\
\hline \hline
1&500,1250 &-0.05,0.09& 47 & 154 & 3.8 & 134 & 143 & 11.2 \\
\hline
2&500,1250 &0.06,0.02& 11 & 123 & 1.0 & 0 & 9 & 0.0 \\
\hline
3&1732,3000 &-0.07,0.04& 7 & 10 & 2.2 & 7 & 8 & 2.5 \\
\hline
4&1732,3000 &0.08,-0.04& 5 & 9 & 1.7 & 6 & 6 & 2.4 \\
\hline
5&1000,1250 &-0.08,0.03 & 108 & 119 &9.9 & 291 & 302 & 16.7\\
\hline
6&1000,1250 &0.07,0.0& 3 & 28 & 0.0 & 15 & 22 & 3.2\\
\hline
\end{tabular}
\end{center}
\caption{The first three columns represent the scenario. The next
three columns give signal and total (including the SM background)
event number for the $Z_1$ production, and the statistical significance
$\sigma =N_{\evt}^{\sig}/\sqrt{N_{\evt}^{\tot}}$ for an integrated
luminosity L=10 fb$^{-1}$. The last three  columns show the same
results for the $Z_2$ production.} \label{tab2}
\end{table}

In presence of two resonances, the four site Higgsless model could
not be easily misidentified. However, in the case depicted in
Fig.~\ref{fig:minv}d where only one resonance survives, the
model-degeneracy problem fully arises. Many models predict in fact
an extra neutral vector boson, one for all the sequential SM (SSM).
Disentangling the various theories, and tracing back the lagrangian
parameters is not an easy task. To this end, a useful observable,
longly studied in the literature, is represented by the
forward-backward charge asymmetry $A_{FB}$. The sensitivity of
$A_{FB}$ measurements to new physics like additional
$Z^\prime$-bosons has been discussed by several authors
\cite{Langacker:1984dc, Dittmar:1996my}. For leptonic Drell-Yan
processes, $A_{FB}$ is defined from the angular distribution with
respect to the quark direction \beq
{d\sigma\over{d\cos\theta^*_l}}\propto {3\over
  8}(1+\cos^2\theta^*_l)+A_{FB}\cos\theta^*_l
\label{eq:angular_dis} \eeq where $\theta^*_l$ is the lepton ($e,\mu$)
angle in the dilepton center-of-mass frame (CM), which can be
derived from the measured four-momenta of the dilepton system in the
laboratory frame. As in $\Pp\Pp$ collisions the original quark
direction is not known, one has to extract it from the kinematics of
the dilepton system. In this analysis, we follow the criteria of
Ref. \cite{Dittmar:1996my} and simulate the quark direction from the
boost of the dilepton system with respect to the beam axis. As a
measure of the boost, we define the dilepton rapidity
$y=1/2\times\ln [(E+p_z)/(E-p_z)]$, and identify the quark direction
through the sign of $y$. We further impose the rapidity to be bigger
than one, $|y|\ge 1$. This cut ensures that the fraction of high
mass dilepton events with a correctly assigned quark direction is
around 80$\%$.

\begin{figure}[]
\begin{center}
\unitlength1.0cm
 \begin{picture}(8,7)
  \xText{-2.4}{5.3}{\small{(a)~~$M_2=1250\GeV, b_{1,2}=(0.07,0)$}}
  \xText{0.3}{0.0}{$\mr{cos}\theta^*_l$}
  \xText{-4.1}{3.}{$L{d\sigma\over{d\cos\theta^*_l}}$}
  \xText{5.7}{5.3}{\small{(b)~~$M_2=1250 \GeV$, $b_{1,2}=(0.07,0)$}}
  \xText{7.3}{0.0}{$\mr{M}_{\inv}(l^+l^-) \normalsize [\GeV]$}
  \xText{4.4}{3.}{$A_{FB}$}
  \xText{0.3}{3.}{$\times 10^{-1}$}
   \put(-4.0,-0.7){\epsfig{file=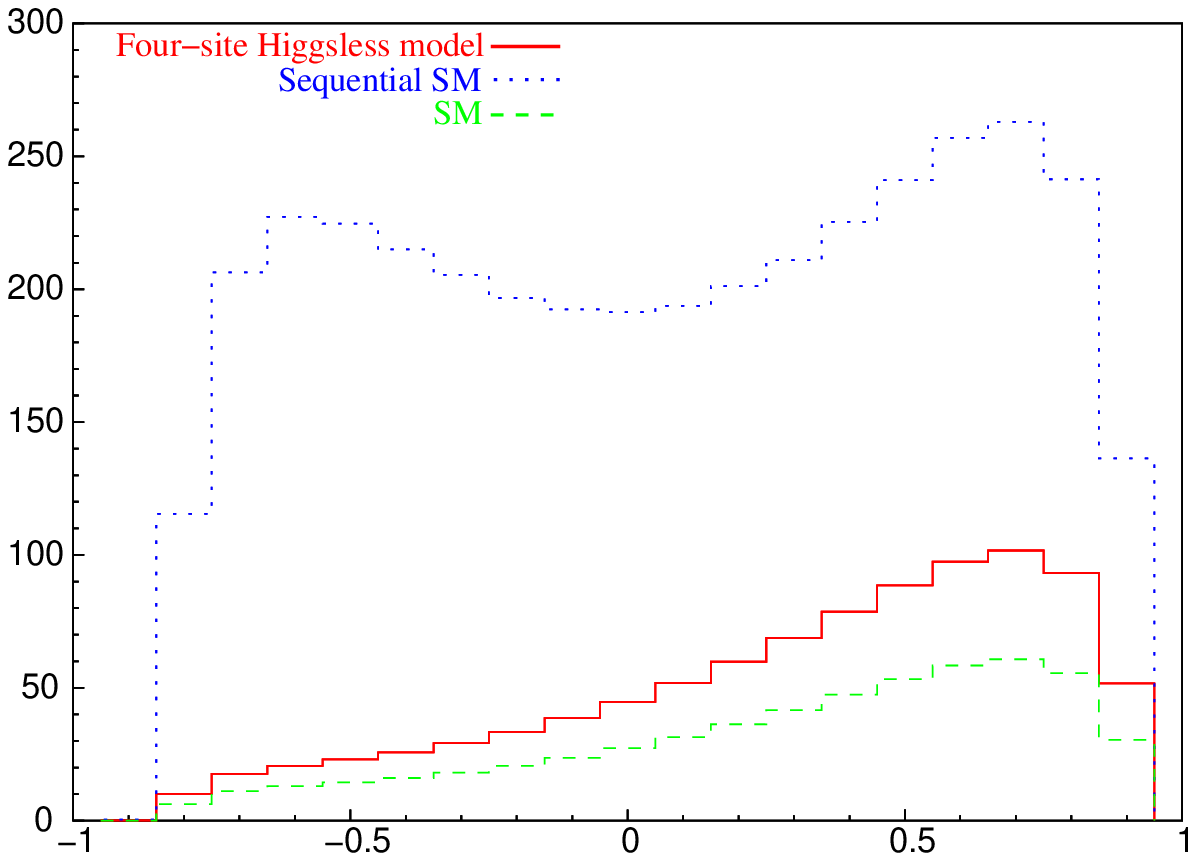,width=12.cm}}
   \put(4.1,-0.7){\epsfig{file=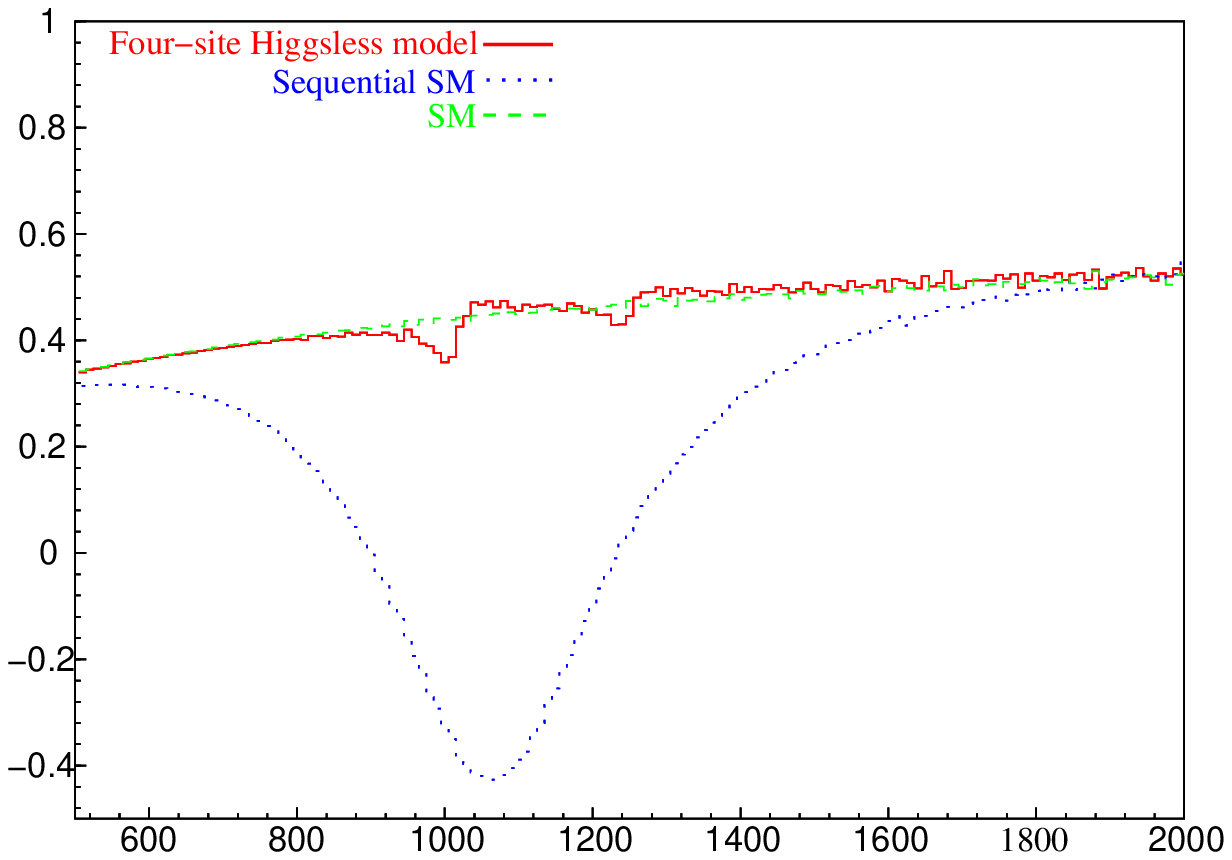,width=12.cm}}
 \end{picture}
\end{center}
\caption{(a) Angular distribution in the lepton angle
with respect to the quark direction in the dilepton rest frame, for the
process $\Pp\Pp\rightarrow l^+l^-$ at the integrated luminosity L=100
fb$^{-1}$. From top to bottom, the three curves describe a sequential-SM
$Z^\prime$-boson with mass $M_2=1250\GeV$, a four site Higgsless $Z_2$-boson
with same mass $M_2=1250\GeV$ and $b_{1,2}=(0.07,0)$, and the SM prediction.
(b) Off-resonance forward-backward asymmetry versus the
dilepton invariant mass, $M_{\inv}(l^+l^-)$, for the process
$\Pp\Pp\rightarrow l^+l^-$ at the integrated luminosity L=100
fb$^{-1}$. From top to bottom, the three curves represent the SM
prediction, a four site Higgsless $Z_2$-boson with mass $M_2=1250
\GeV$ and $b_{1,2}=(0.07,0)$, and a sequential-SM $Z^\prime$-boson with mass
$M=1250\GeV$. In both figures, we sum over $e,\mu$. Standard cuts and legends
as in the text.} \label{fig:asymmetry}
\end{figure}

We study both on-resonance and off-resonance asymmetries for the
single-resonant scenario depicted in Fig.~\ref{fig:minv}d. In the
first case, we consider the on-peak region
$|M_{\inv}(l^+l^-)-M_2|<3\Gamma_2$, and plot the angular
distribution of Eq. (\ref{eq:angular_dis}), including signal and
background. In order to illustrate how $A_{FB}$ could help in
solving the model-degeneracy problem, in Fig.~\ref{fig:asymmetry}a
we compare the four site Higgsless model with the sequential SM,
which both predict an extra neutral vector boson ($Z_2$). The SM
result is given as a reference. As one can see, the two models
present quite a different shape for the differential cross section
in $\cos\theta^*_l$. The forward-backward asymmetry is much more
pronounced in the four site Higgsless model, owing to the sensible
difference between left and right-handed fermion-boson couplings, as
reported in Tab.~\ref{tab4} of Appendix \ref{numbers}. The
coefficient, $A_{FB}$, in Eq. (\ref{eq:angular_dis}) is indeed
proportional to the difference between the two couplings squared,
$A_{FB}\propto [(a^f_{2L})^2-(a^f_{2R})^2] $.

As a further tool to separate the two models, in Fig.~\ref{fig:asymmetry}b we
plot the asymmetry for continuum dilepton events at high CM energy scales.
Here, we integrate over the lepton angle in the forward and backward
region, separately, and plot the difference between the resulting
forward and backward differential cross sections in the dilepton
invariant mass, normalized to their sum \beq A_{FB}= \left
[{d\sigma^F\over{d\M_{\inv}}}-{d\sigma^B\over{d\M_{\inv}}}\right ]/
\left
[{d\sigma^F\over{d\M_{\inv}}}+{d\sigma^B\over{d\M_{\inv}}}\right ].
\eeq
The SM prediction is shown as a reference. The measurement of
the dips, pointing at different CM-energy values according to the
considered model, represents an additional  powerful tool to
understand the nature of the neutral resonance. In this particular case,
the off-resonance $A_{FB}$ could also reveal the double-resonant
structure of the four site Higgsless model signal, not appreciable in the
dilepton invariant mass distribution (compare with Fig.~\ref{fig:minv}d).
The first dip, related to the $Z_1$-boson mass, gives indeed a 3.8$\sigma$
effect if compared to the SM prediction, for an high luminosity
L=100$\fba^{-1}$.


\begin{figure}[]
\begin{center}
\unitlength1.0cm
 \begin{picture}(8,7)
 \put(-3.8,0.4){\epsfig{file=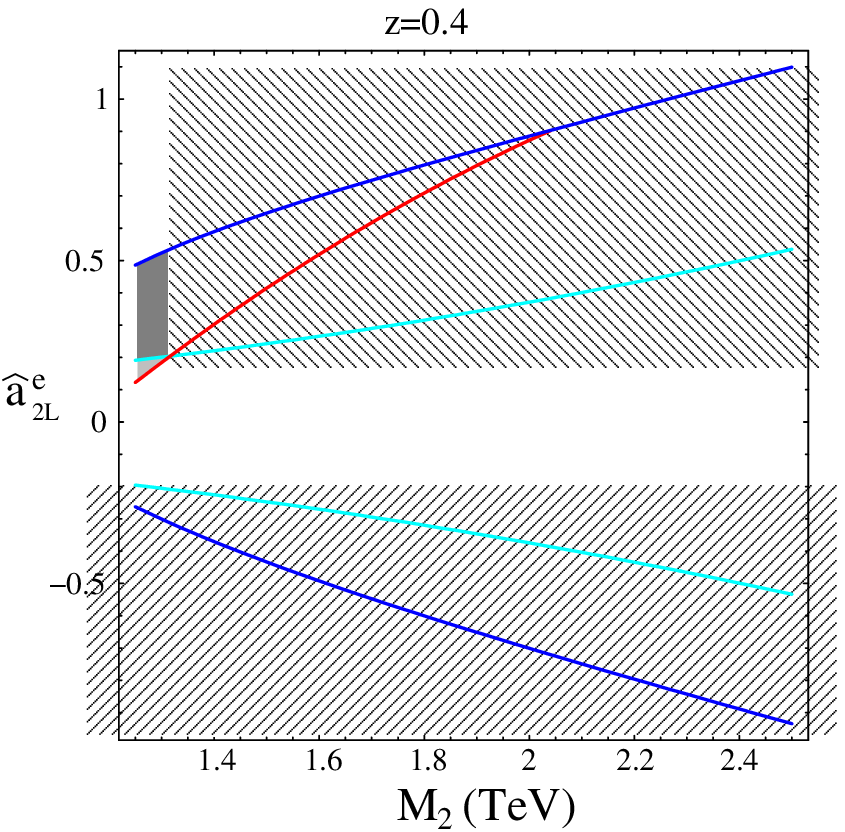,width=7.cm}}
  \put(4.5,0.4){\epsfig{file=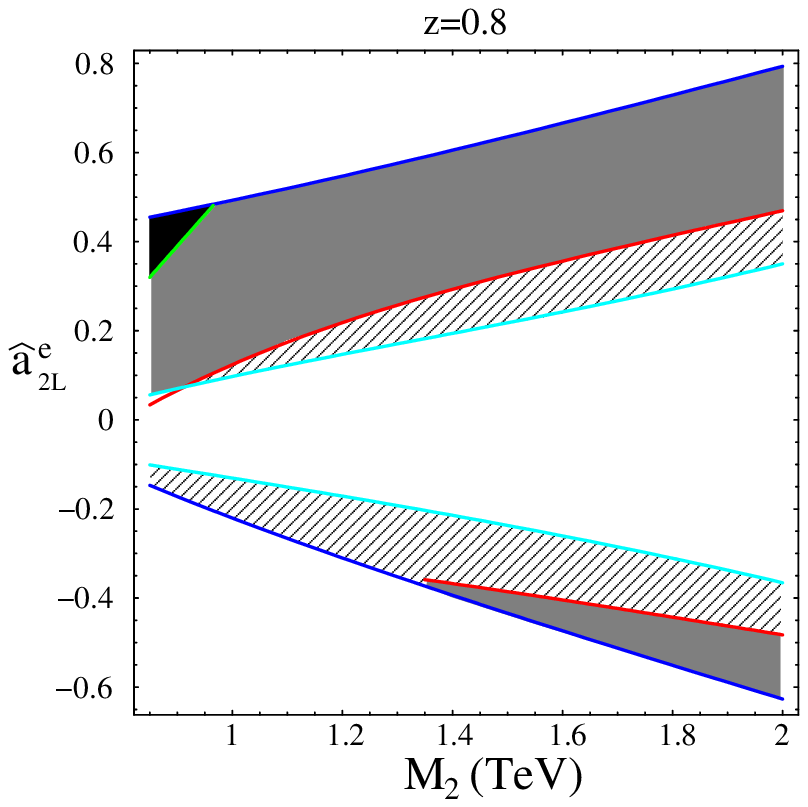,width=7.cm}}
 \end{picture}
\end{center}
\caption{5$\sigma$-discovery plot at L=100$\fba^{-1}$ in the plane
 $({\hat a}^e_{2L}, M_2)$  (in the coupling the electric charge -$e$ is
 factorized). The upper and lower parts are excluded by EWPT, the
black triangle in the right panel is the region excluded by the
direct search at the Tevatron. Inside the dark-grey regions both
$Z_{1,2}$ are visible; inside the grey (dashed) ones only $Z_1$
($Z_2$) can be detected. Inside the central uncolored region no
resonance is visible in the Drell-Yan channel. Left (right) panel:
$z=0.4$ ($z=0.8$).} \label{fig:visibilita}
\end{figure}

After analyzing the spectrum of the extra $Z_{1,2}$-bosons, and how
their nature could be investigated through asymmetry measurements,
let us summarize the possibility to detect these new particles at
the LHC as a function to the integrated luminosity.
\begin{figure}[t]
\begin{center}
\unitlength1.0cm
 \begin{picture}(8,7)
  \xText{-0.0}{1.7}{\small{$(z=0.4)$}}
  \xText{5.1}{1.7}{\small{$(z=0.4)$}}
  \xText{-0.0}{-5.3}{\small{$(z=0.8)$}}
  \xText{8.4}{-5.3}{\small{$(z=0.8)$}}
  \put(-4.2,0.5){\epsfig{file=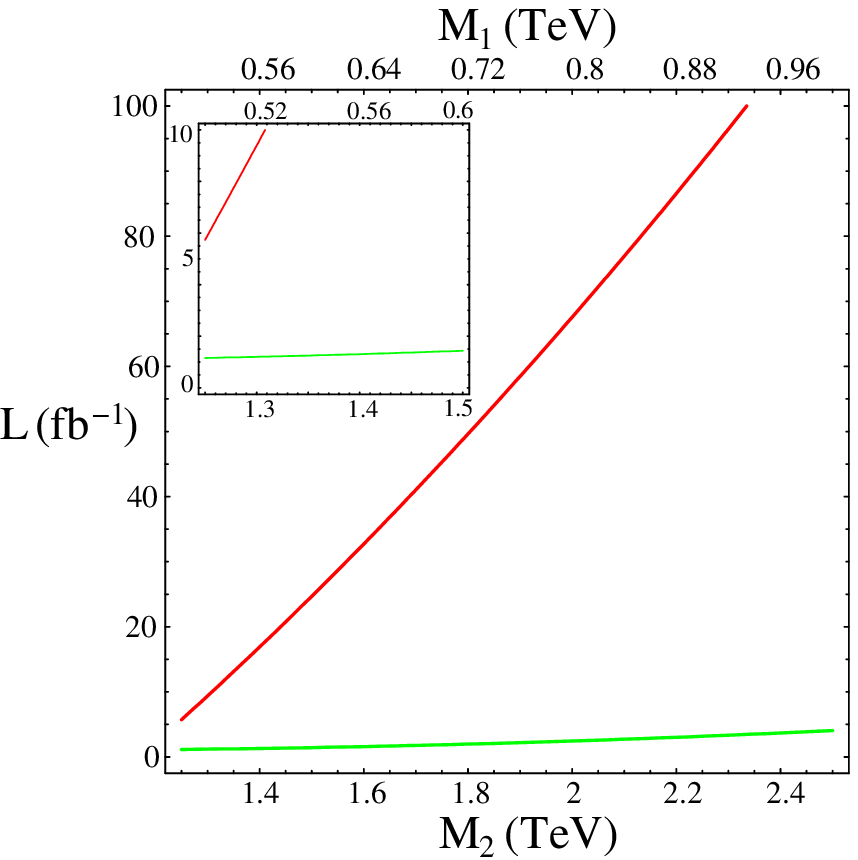,width=6.5cm}}
  \put(3.6,0.5){\epsfig{file=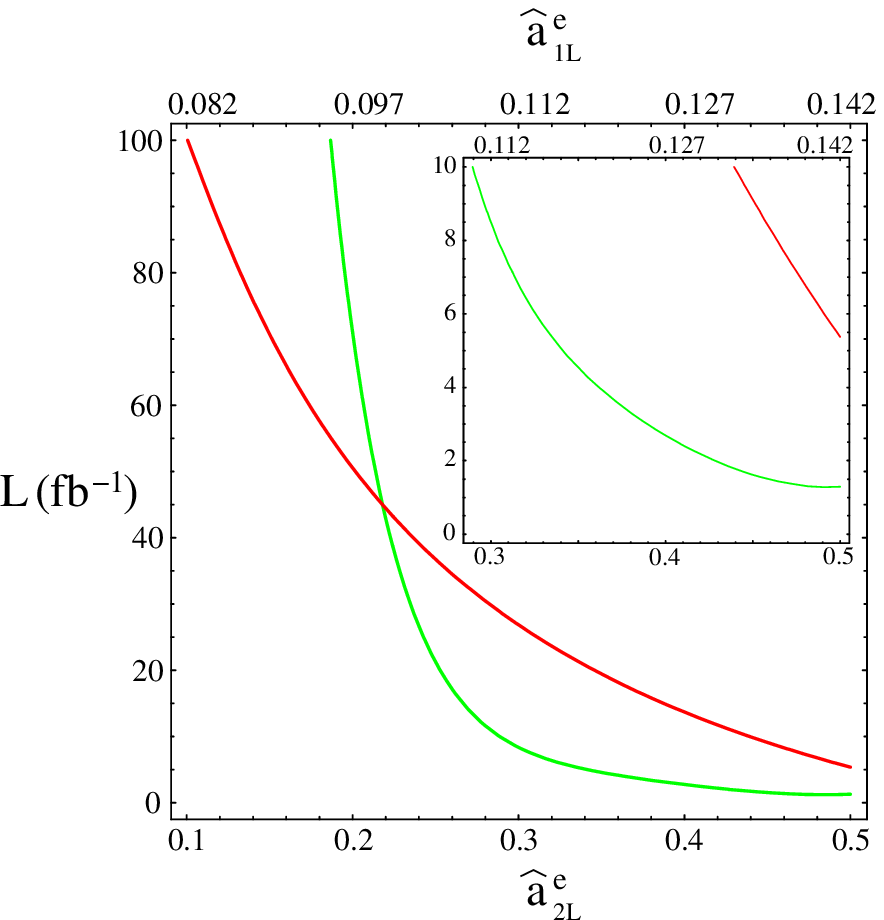,width=6.5cm}}
  \put(-4.2,-6.5){\epsfig{file=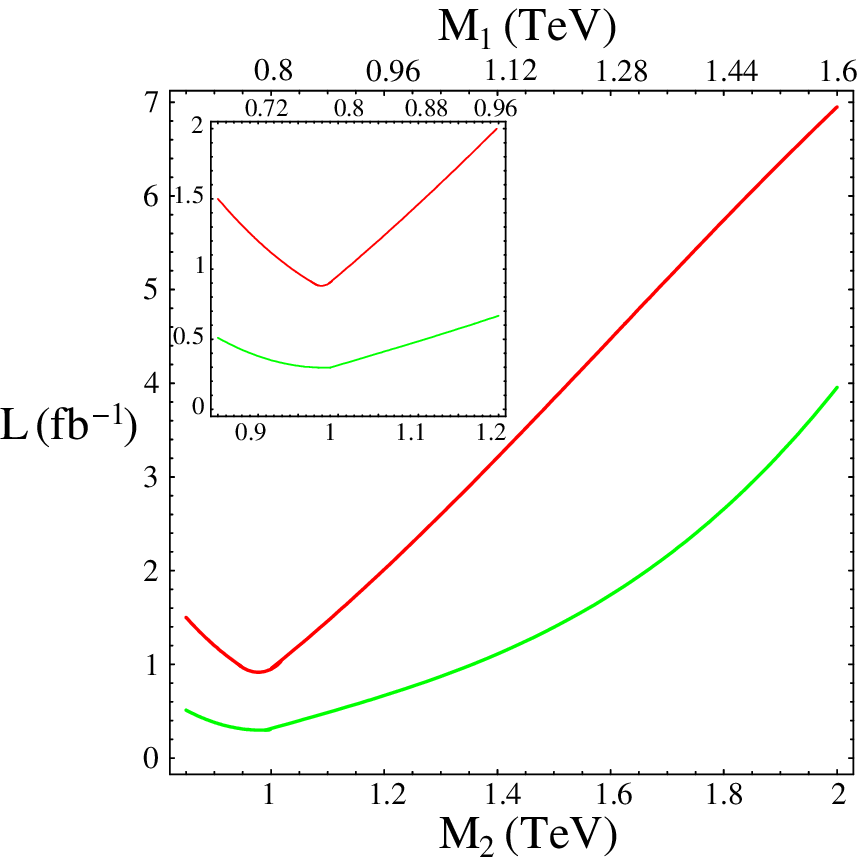,width=6.5cm}}
  \put(3.6,-6.5){\epsfig{file=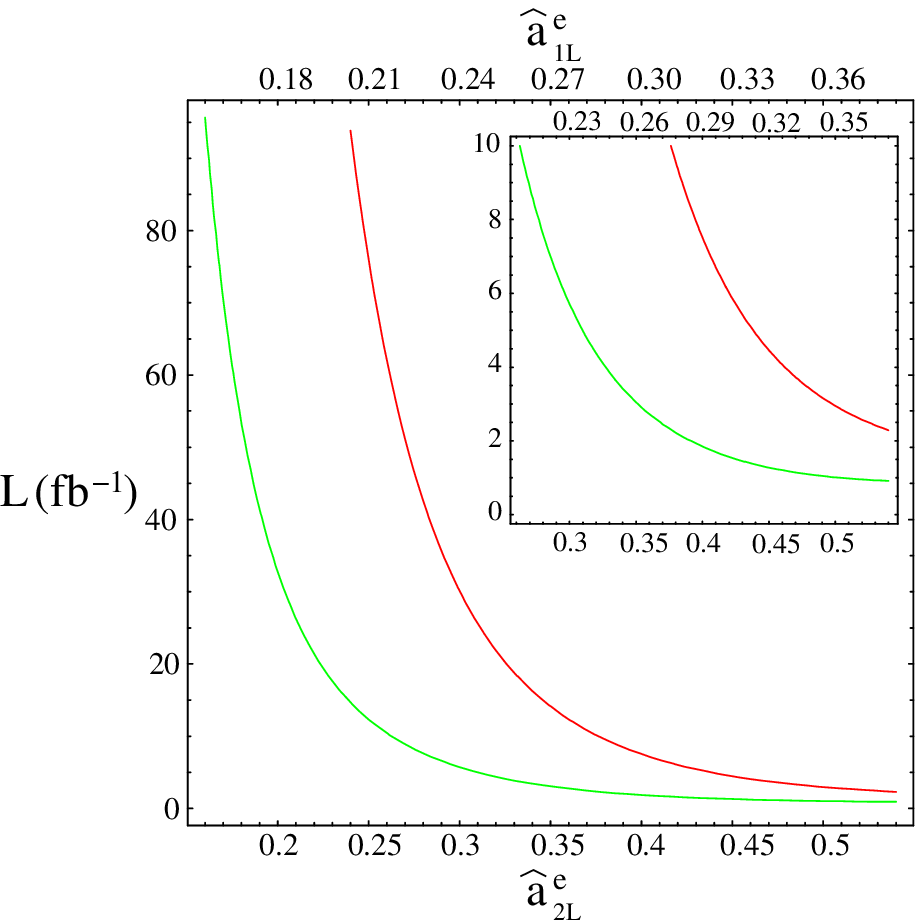,width=6.5cm}}
 \end{picture}
\end{center}
\vskip 6.cm \caption{Left panels: Luminosity vs $Z_2$ mass needed
for a 5$\sigma$-discovery for $z=0.4$ (up) and $z=0.8$ (down). We
assume the maximum value for the fermion-boson couplings allowed by
EWPT and Tevatron, as shown by the upper curves in
Fig.~\ref{fig:visibilita}. The discovery curves are for $Z_1$ (red,
darker) and $Z_2$ (green, lighter) gauge bosons. Right panels:
Luminosity vs electron-boson left-handed coupling (the electric
charge -$e$ is factorized) needed for a 5$\sigma$-discovery for
 $z=0.4$ (up) and $z=0.8$ (down) with $M_2=1250~GeV$ and $M_1=z M_2$.
 The discovery curves are for $Z_1$ (red,
darker) and $Z_2$ (green, lighter) gauge bosons. In all the figures,
we sum over $e,\mu$ and apply standard cuts. The inset plots show
the zoomed-in low-luminosity region.} \label{fig:lumi}
\end{figure}
In Fig.~\ref{fig:visibilita} we plot the 5$\sigma$-discovery
contours at L=100 fb$^{-1}$ in the plane $({\hat a}_{2L}^e, M_2)$,
where ${\hat a}_{2L}^e$ is the left-handed coupling between the
$Z_2$-boson and the SM electron (in -$e$ units) and $M_2$ is related
to the $Z_2$-mass by Eq. (\ref{eq:M2}). We have considered $M_2\leq
2(2.5)$ TeV for $z=0.8(0.4)$ in order to agree with the strongest
partial wave unitarity bound shown in Fig. \ref{a0} and two
different values of the parameter $z$. The coupling $a^e_{1L}$ is
fixed by using
 the expression of $\eps_3$
given in Eq. (\ref{epsilon3}) after including  radiative
corrections, and comparing it to the experimental value,
$\eps_3^{exp}$. Looking at the upper part of the plots and going
from top to bottom, the first curve represents the ${\hat a}_{2L}^e$
maximum values allowed by EWPT as a function of $M_2$. The small
black triangle in the right panel gives the direct limits from
Tevatron for a luminosity $L=4\fba^{-1}$ \cite{tevatron-lumi}. This
region is not visible in the left panel because of the different
mass range. As to the detection, the dark-grey region between the
first two curves represents the parameter space where both $Z_{1,2}$
resonances are simultaneously visible in the Drell-Yan channel,
while the dashed region shows the range where only the $Z_2$-boson
could be detected. The small grey region for low masses shows the
range where only the $Z_1$-boson could be detected. The lower part
of the plot is specular. In the central uncolored region one should
measure other channels like di-boson production and $WW$ scattering.
Notice that for low $z$ values, the $Z_1$-boson couplings to
fermions are much more bounded by EWPT (Eq.~(\ref{epsilon3})), as a
consequence the $Z_1$ visibility region by DY production is smaller
with respect to larger $z$ values.

The main information one gets from Fig.~\ref{fig:visibilita} is that
the four site Higgsless model can be explored at the LHC in the
favoured Drell-Yan channel, over a large portion of the parameter
space, without invoking much more complex multi-particle processes
like di-boson production or vector boson fusion as in the usual
Higgsless literature \cite{Belyaev:2007ss, He:2007ge}. Moreover, the
new $Z_{1,2}$ gauge bosons could be discovered in the early stage of
the LHC data taking at very low luminosity.

Fig. \ref{fig:lumi} (left panels) shows in fact that the minimum
integrated luminosity required to observe the low edge of the
spectrum is around L$\simeq$ 1-2${\rm fb}^{-1}$. Notice that the
raise of luminosity for low masses for $z=0.8$ is due to the
Tevatron exclusion region shown in Fig. \ref{fig:visibilita}. In
Fig. \ref{fig:lumi} (right panels), we plot in addition the
luminosity needed for a 5$\sigma$ discovery of a $Z_2$-boson with
mass $M_2=1.25\TeV$ and a $Z_1$-boson with mass $M_1=z M_2$  as a
function of the left-handed electron-boson coupling.

\subsubsection{$W_1^\pm$ and $W_2^\pm$ production at the LHC}
\label{se:charged_production}

In this section, we present some cross sections and distributions
for the leptonic process $\Pp\Pp\to l\nu_l$ with $l=\Pe,\mu$ and
$l\nu_l=l^-\bar\nu_l, l^+\nu_l$. These
final states allow to analyze the production of the four charged
extra gauge bosons, $W^\pm_{1,2}$, predicted by the four site
Higgsless model.

As we said previously, we are interested in the high energy region where
the new strongly interacting vector bosons are expected to be produced.
We thus impose an additional cut on the transverse momentum of the lepton
pair, $P_T(l\nu_l)\ge 150\GeV$, which selects large CM-energies
($\sqrt{s}\ge 300\GeV$).
\begin{figure}[t]
\begin{center}
\unitlength1.0cm
 \begin{picture}(8,7)
  \xText{-2.5}{5.3}{\small{(a)~~$M_{1,2}=(500, 1250)\GeV$}}
  \xText{-1.3}{-0.1}{$\mr{M}_t(l\nu_l) \normalsize [\GeV]$}
  \xText{-4.}{3.}{$N_{evt}$}
  \xText{5.2}{5.3}{\small{(b)~~$M_{1,2}=(1732, 3000)\GeV$}}
  \xText{6.7}{-0.1}{$\mr{M}_t(l\nu_l) \normalsize [\GeV]$}
  \xText{3.9}{3.}{$N_{evt}$}
  \xText{-2.5}{-0.85}{\small{(c)~~$M_{1,2}=(1000, 1250)\GeV$}}
  \xText{-1.3}{-6.3}{$\mr{M}_t(l\nu_l) \normalsize [\GeV]$}
  \xText{-4.}{-3.3}{$N_{evt}$}
  \xText{5.2}{-0.85}{\small{(d)~~$M_{1,2}=(1000, 1250)\GeV$}}
  \xText{6.7}{-6.3}{$\mr{M}_t(l\nu_l) \normalsize [\GeV]$}
  \xText{3.9}{-3.3}{$N_{evt}$}
  \put(-4.2,-0.7){\epsfig{file=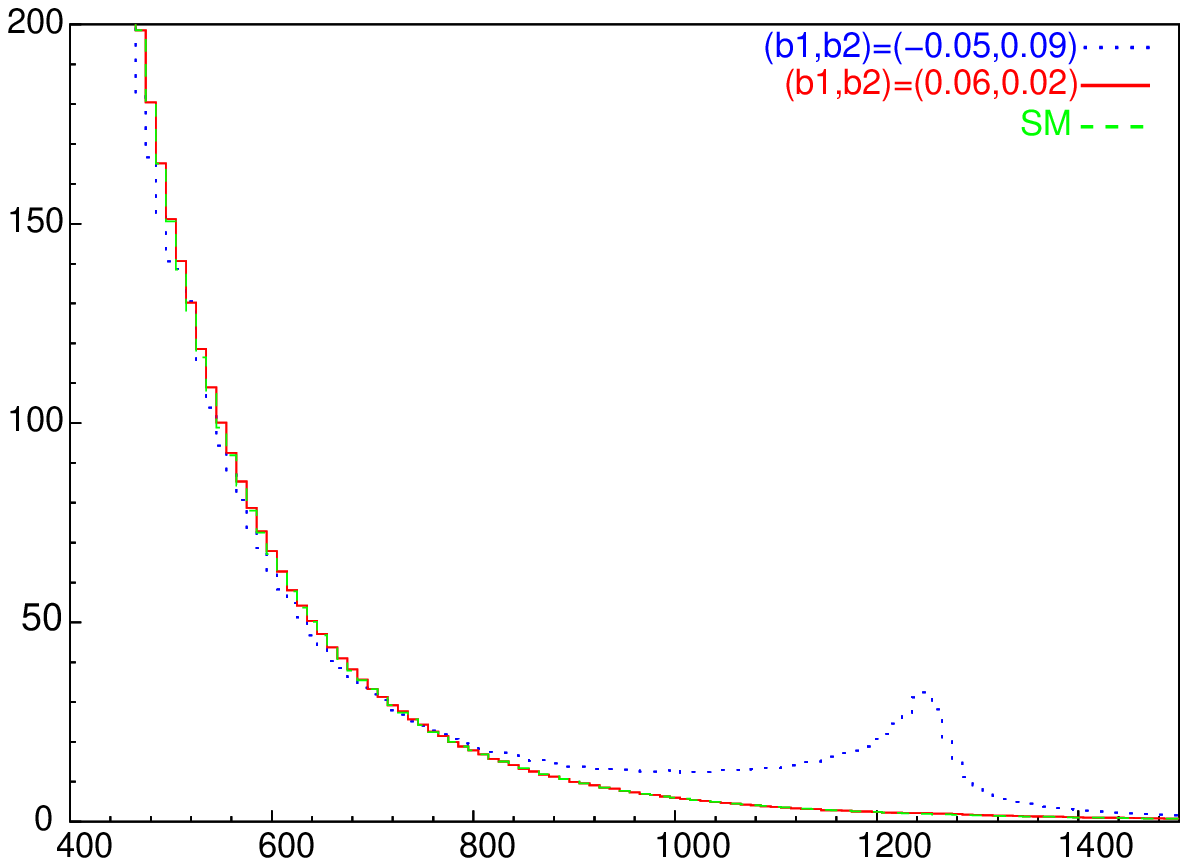,width=12.cm}}
  \put(3.6,-0.7){\epsfig{file=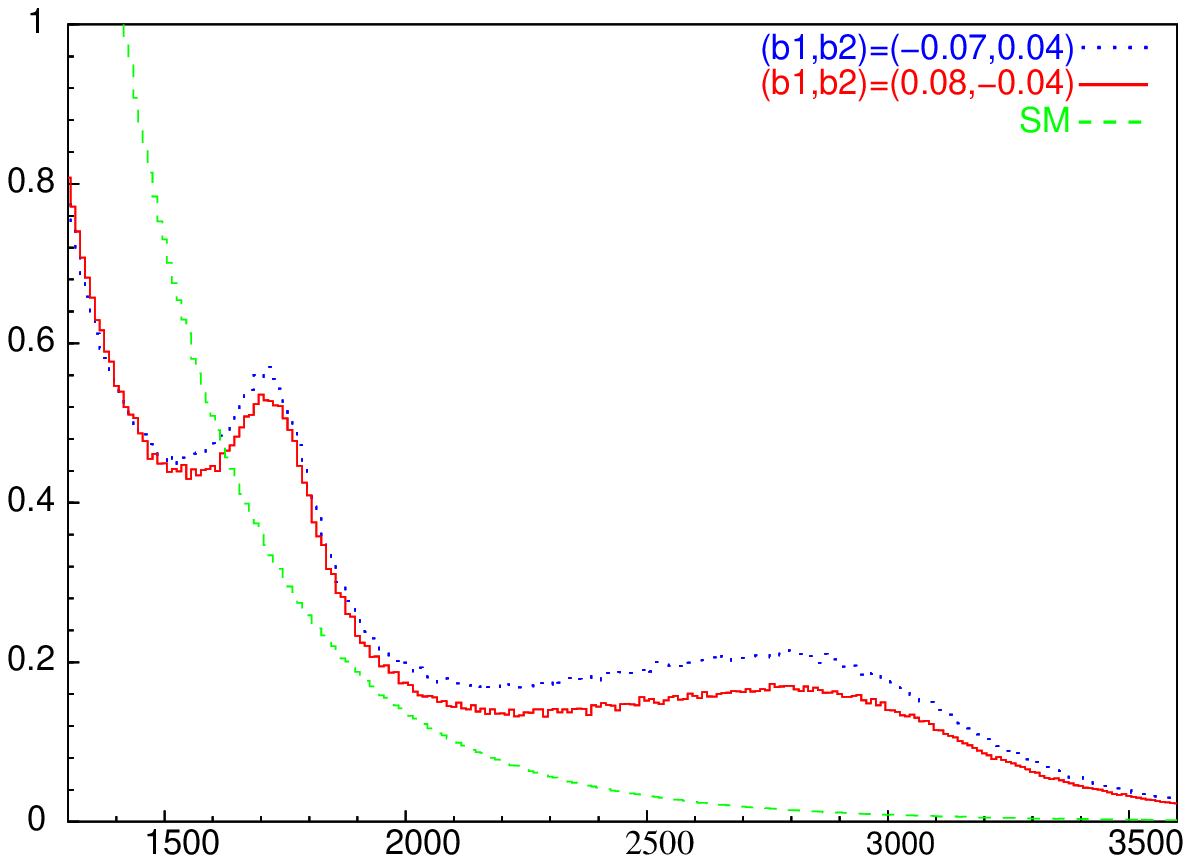,width=12.cm}}
  \put(-4.2,-6.9){\epsfig{file=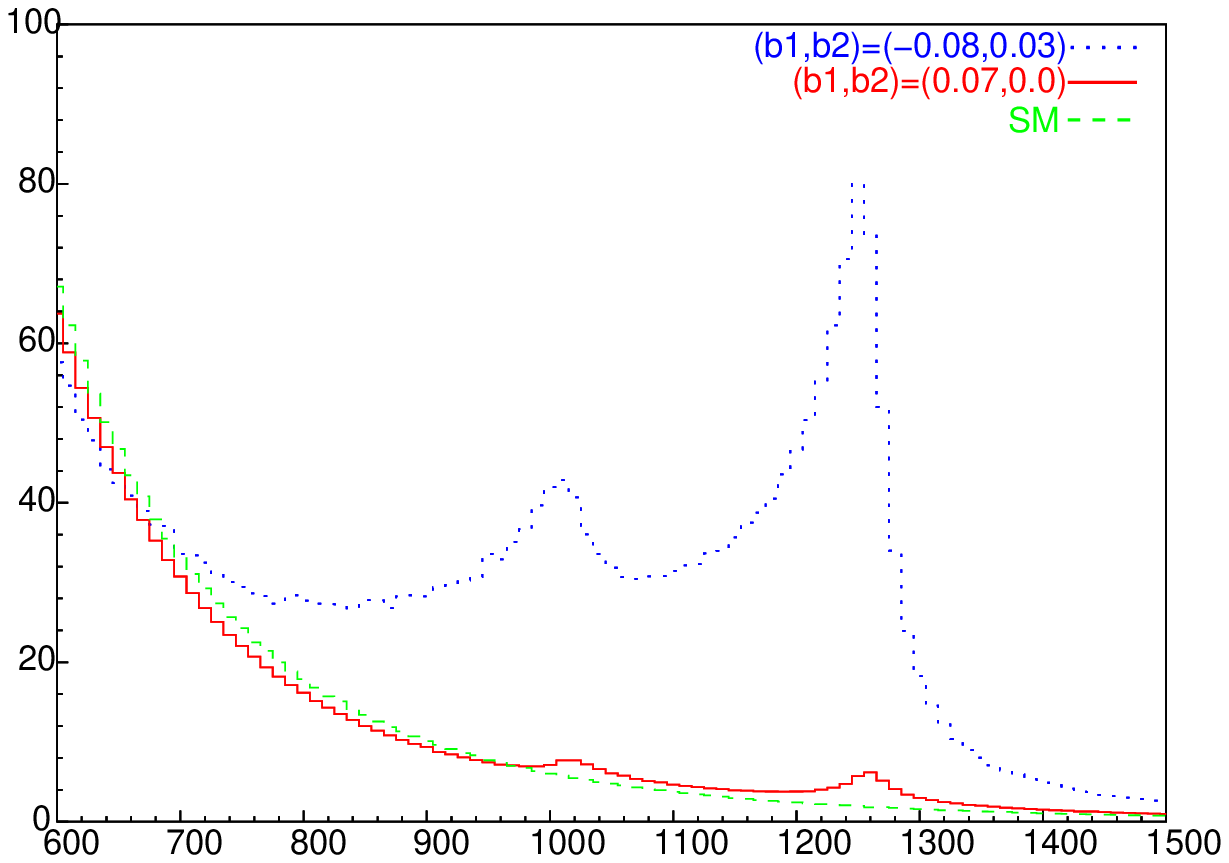,width=12.cm}}
  \put(3.6,-6.9){\epsfig{file=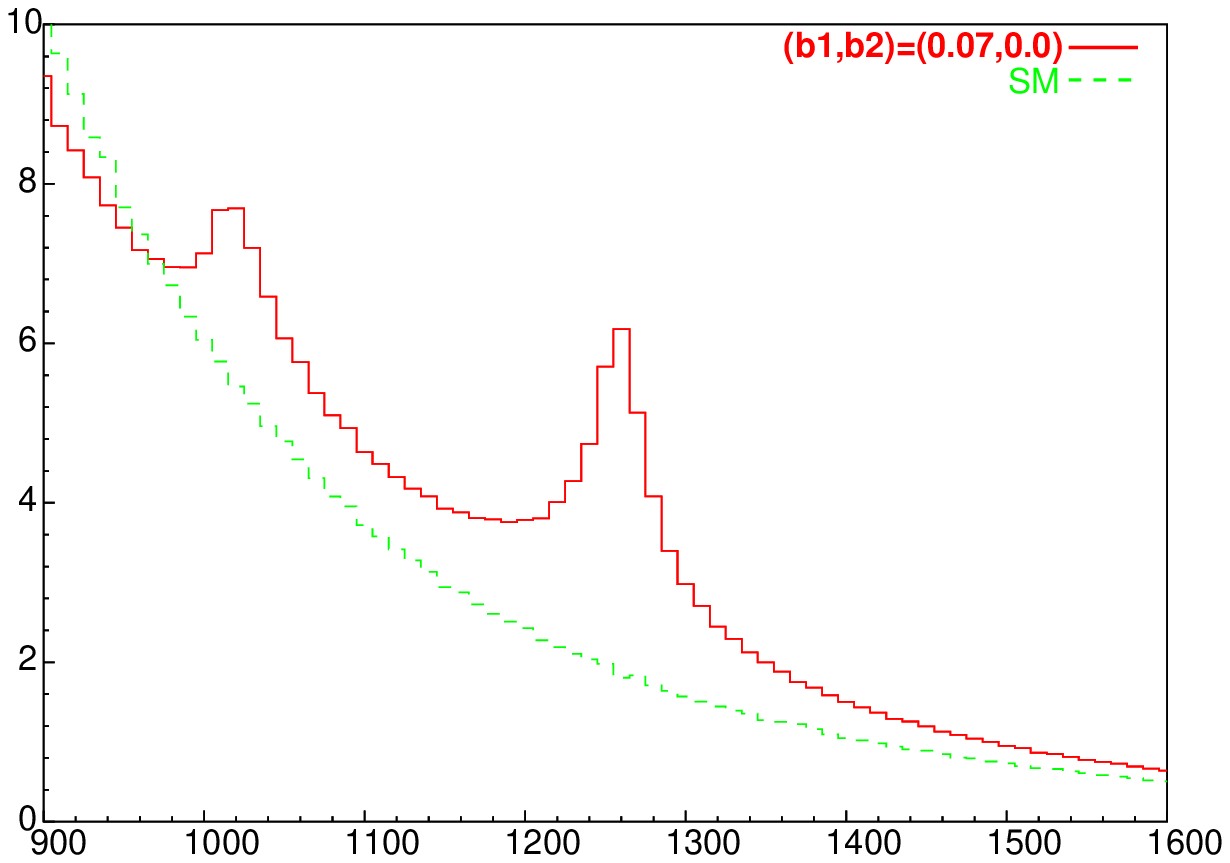,width=12.cm}}
 \end{picture}
\end{center}
\vskip 6.cm \caption{Total number of events in a 10$\GeV$-bin versus
the lepton transverse mass, $M_t(l\nu_l)$, for the process
$\Pp\Pp\rightarrow l\nu_l$ at the integrated luminosity $L=10$
fb$^{-1}$ for the six scenarios of Table \ref{tab:scenarios}. We sum over $e,\mu$ and charge conjugate channels. Standard cuts
and legends as in the text.}
\label{fig:mt}
\end{figure}
In order to illustrate spectrum and behaviour of the new charged
particles at the LHC, we have chosen to analyze the distribution in the
transverse mass of the lepton pair, $M_t(l\nu_l)$,
for the six scenarios of Tab.~\ref{tab:scenarios}, that is for
three values of the $z$-parameter, $z=(0.4, 1/ \sqrt{3}, 0.8)$, and
various $b_{1,2}$-sets. The corresponding charged fermion-boson
couplings are summarized in Tab.~\ref{tab1} of Appendix~\ref{numbers}.

In analogy with the neutral case, in Fig.~\ref{fig:mt} we plot the total
number of events as a function of the dilepton transverse mass, $M_t(l\nu_l)$,
for the six aforementioned scenarios. From top to bottom, the three curves in
each plot represent the first $b_{1,2}$-setup, the latter, and the
SM prediction at fixed $M_{1,2}$ mass parameter. We sum over $e,\mu$ and
charge conjugate processes. We moreover apply standard acceptance cuts.

These examples show again some peculiar predictions of the model.
First of all, the spectra of neutral and charged gauge sectors are
almost degenerate ($M_{i,c}\simeq M_{i,n}$ with $i=1,2$). The same
peaking structure is thus expected in both neutral and charged
Drell-Yan channels. This is clearly visible in Figs. \ref{fig:mt}b,c
where the peaks in the lepton transverse mass are centered on the
same values as those ones in the dilepton invariant mass reported in
Figs.~\ref{fig:minv}b,c. In these cases, we have a cross-checkable
double signal. The situation can however be different, depending on
masses and couplings. For example, if one considers the first
scenario of Tab. \ref{tab:scenarios}, one realizes that while in the
neutral channel the two new resonances are both visible (at least at
parton level), in the charged channel only the heavier one survives.
The first $W_1^\pm$-boson peak is indeed washed out by the very low
couplings between charged extra gauge bosons and SM fermions (see
Tab. \ref{tab1} in Appendix~\ref{numbers}). In this case, in order
to have the full information, exploring the two channels is
mandatory. The last plot in Fig. \ref{fig:mt}d is again a mirror of
Fig. \ref{fig:minv}d, i.e. in this scenario only the heavier
resonances $W_2^\pm$ and $Z_2$ could be detected in their respective
channels (with $L\sim 100~{\rm fb}^{-1}$).

\begin{table}
\begin{center}
\begin{tabular}{|c|c|c|c|c|c|c|c|c|c|}
\hline &$M_{1,2} (\GeV)$ & $b_{1,2}$ &$M_t^{cut}(\GeV)$ &$N_{\evt}^{\sig}(W_1)$ &
$N_{\evt}^{\tot}(W_1)$ & $\sigma (W_1)$ & $N_{\evt}^{\sig}(W_2)$ &
$N_{\evt}^{\tot}(W_2)$
& $\sigma (W_2)$ \\
\hline \hline
1&500,1250 &-0.05,0.09&400 &36 & 2435 & 0.7 & 776 & 2214 & 16.5 \\
\hline
2&500,1250 &0.06,0.02&400 &0 & 2609 & 0 & 1 & 1807 & 0 \\
\hline
3&1732,3000 &-0.07,0.04&1500 &10 & 18 & 2.4 & 24 & 26 & 4.7 \\
\hline
4&1732,3000 &0.08,-0.04&1500 &9 & 14 & 2.4 & 22 & 24 & 4.5 \\
\hline
5&1000,1250 &-0.08,0.03 &700 &808 & 1230 &23.0 & 1112 & 1189 & 32.3\\
\hline
6&1000,1250 &0.07,0.0&700 &12& 443 & 0.6 & 17 & 88 & 1.8\\
\hline
\end{tabular}
\end{center}
\caption{The first three columns represent the scenario. The fourth one
shows the cut on the dilepton transverse mass $M_t(l\nu_l)$. The next
three columns give signal and total (including the SM background)
event number for the $W^\pm_{1}$ production, and the statistical
significance $\sigma =N_{\evt}^{\sig}/\sqrt{N_{\evt}^{\tot}}$ for an integrated
luminosity L=10 fb$^{-1}$. The last three columns give the corresponding results for $W_2^\pm$ production.} \label{tab:nevt_charged}
\end{table}

In order to estimate the detection rate expected at the LHC for the
Drell-Yan production of the extra $W_{1,2}^\pm$ gauge bosons, in
Tab.~\ref{tab:nevt_charged} we have listed signal and total event
number for the six scenarios given in Tab.~\ref{tab:scenarios}. We
have evaluated the number of $W^\pm_{1,2}$ events in the kinematical
region defined by the cut: $M_t(l\nu_l)\ge M_t^{cut}$.
The value of $M_t^{cut}$ is chosen as the value where the total
number of events is equal to the SM background. The number of events
 is obtained by integrating in transverse mass between $M_t^{cut}$
and $M_1+\Gamma_1$ for the first resonance and between
$M_1+\Gamma_1$ and $M_2+\Gamma_2$ for the second one.
\begin{figure}[t]
\begin{center}
\unitlength1.0cm
 \begin{picture}(8,7)
 \put(-3.8,0.4){\epsfig{file=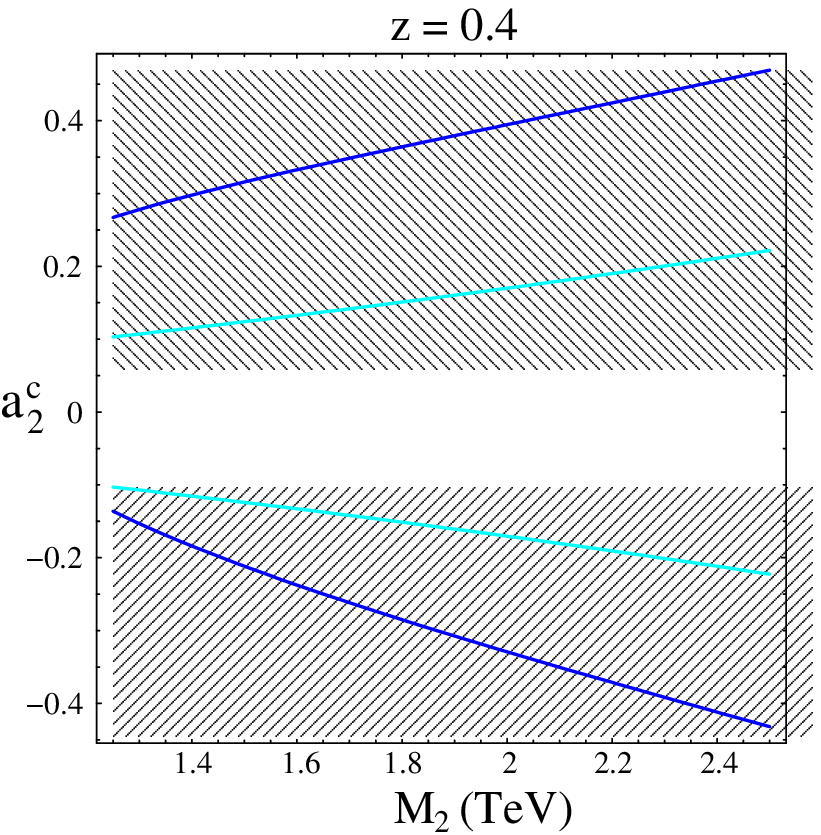,width=6.8cm}}
  \put(4.5,0.4){\epsfig{file=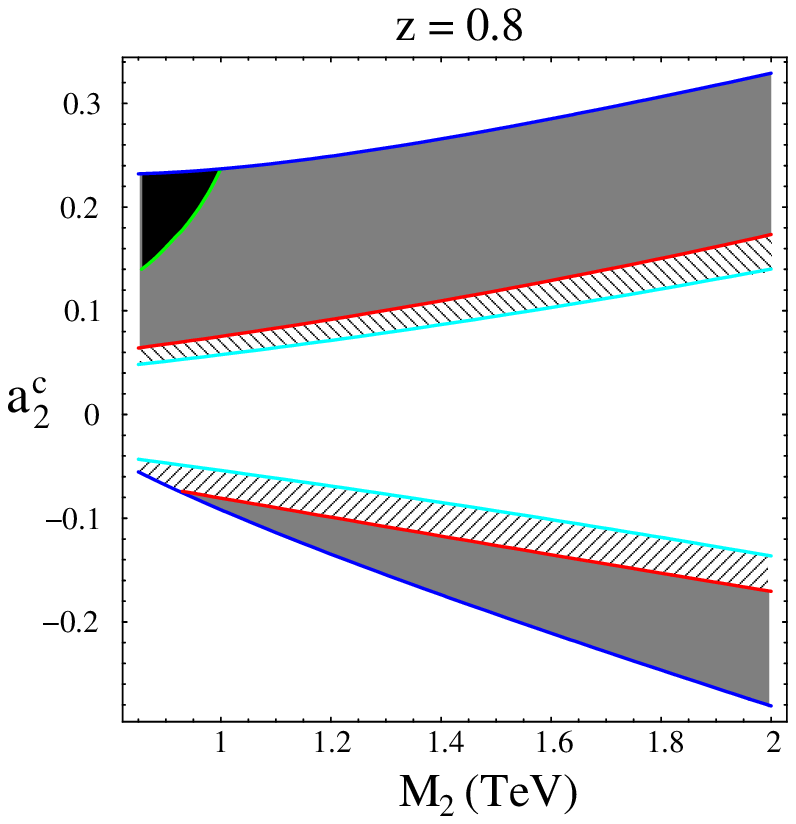,width=7.0cm}}
 \end{picture}
\end{center}
\caption{5$\sigma$-discovery plot at L=100$\fba^{-1}$ in the plane
 $(a^c_{2}, M_2)$. The upper and lower parts are excluded by EWPT, the
black triangle is the region excluded by the direct search at the
Tevatron. Inside the grey regions both $W_{1,2}$ are visible; inside
the dashed ones only $W_2$ can be detected. Inside the central
uncolored region no resonance is visible in the Drell-Yan channel.
Left (right) panel: $z=0.4$  ($z=0.8$)} \label{fig:visibilitac}
\end{figure}
>From Tab.~\ref{tab:nevt_charged}, one can see that the charged
Drell-Yan channel has generally a higher sensitivity compared to the
neutral one, for a given scenario. In some cases the statistical
significance is about a factor two bigger than the previous neutral
case, however the charged Drell-Yan observables are not that clean.
In order to have a well defined information on the four site
Higgsless model predictions, neutral and charged Drell-Yan channels
are thus complementary. And, more important, there are regions in
the parameter space where they could be both investigated for the
search of all six extra gauge bosons, $W_{1,2}^\pm$ and $Z_{1,2}$,
at the LHC start-up with a luminosity of the order of L$\simeq$ 1-2
fb$^{-1}$ for $M_{1,2}\le$ 1 TeV.

Let us finally comment  on the dependence of the discovery potential in the charged channel
 in the parameter space of the model.
In Fig.~\ref{fig:visibilitac} we plot the 5$\sigma$-discovery contours at L=100
 fb$^{-1}$ for $z=0.4$ (left) and $z=0.8$ (right) in the plane $(a^c_2,M_2)$ where $a_2^c$ is the $W_2$
charged coupling and $M_2$  is related to the $W_2$ mass by Eq.
(\ref{M2c}). The upper and lower parts of the plot are excluded by
EWPT, the black triangle is the region excluded by the direct search
at the Tevatron with 4 fb$^{-1}$. Inside the dark-grey regions both
$W_{1,2}$ are visible; inside the dashed ones only $W_2$ can be
detected. Inside the central uncolored region no resonance is
visible in the Drell-Yan channel. Notice that, while for $z=0.8$ the
region in which one looses the first  resonance is small, for
smaller $z$-values this region increases and for example, for
$z=0.4$, the four site model could be misidentified with schemes
with only one  triplet of new resonances. For this portion of
parameter space the di-boson and/or the fusion channels are
mandatory for disentangling among different models.

\section{Conclusions}

In this work we have studied the phenomenological consequences of the four
site Higgsless model, a deconstructed theory which can derive from the
discretization of the fifth dimension on a lattice, and is based on the
$SU(2)_L\times SU(2)_1\times SU(2)_2\times U(1)_Y$ gauge symmetry. The model
represents an
extension of the minimal three site version (or BESS model), longly
investigated in the literature, which includes three heavy vector bosons,
$W^\pm_1$ and $Z_1$. The four site model extends the gauge sector to four
charged and two neutral extra gauge bosons, $W^\pm_{1,2}$ and $Z_{1,2}$. We
have analysed their properties and the prospects for their direct search at
the LHC.

One of the strong motivations for Higgsless models is their ability
to delay the unitarity violation of VBS amplitudes to energy scales
higher than those predicted by the SM without a light Higgs, just
via the exchange of the above mentioned extra gauge bosons. The
drawback is the strong tension between the unitarity requirement and
the bounds coming from the electroweak precision tests. In the
minimal three site model, the request to reconcile unitarity and
EWPT brings to a fermiophobic scenario, where the extra gauge bosons
can barely interact with ordinary matter. In this case, the only
production processes available to search for the new particles are
those dominated by boson-boson interactions. Hence, one has to rely
on difficult multi-particle processes which require high luminosity
to be revealed, that is vector boson fusion and triple gauge boson
production. The recent Higgsless literature has been focused on this
side.

The novelty of the four site Higgsless model consists in reconciling
unitarity and EWPT bounds without imposing the extra vector bosons
to be fermiophobic, owing to the inclusion of direct fermion-boson
couplings in addition to those ones coming from usual mixing terms.
We have analysed in detail the constraints on masses and couplings
of the extra gauge bosons coming from unitarity requirement and EW
precision data consistency. We have found that, asking for all VBS
amplitudes with both SM and extra gauge bosons as external states to
be unitarized, the perturbative regime can be extended up to energy
scales of the order of $\sqrt{s}\simeq 3\TeV$, thus higher than
those predicted by the SM with no light Higgs and by the three site
minimal version. Hence, the spectrum of the new gauge bosons must
lie within a few TeV, a region which represents the main effective
energy range available at the LHC. We have moreover investigated the
impact of the EW precision measurements expressed in terms of the
$\eps_i$ parameters on the couplings of the extra gauge bosons to
ordinary matter. While $\eps_3$ gives the strongest bound, imposing
a strict relation between the couplings of the two vector boson
triplets, $\eps_1$ weakly limits their size owing to the SU(2)
custodial symmetry ($\eps_2$ is uneffective). The fermion-boson
couplings can be thus of the same order of the SM ones. As a
consequence, the Drell-Yan process becomes an open channel for the
direct search of the extra gauge bosons at the LHC.

We have analysed in detail
the potential detection rate of the new particles, evaluating cross sections
and distributions for charged and neutral Drell-Yan processes at the LHC. In
order to show a full view, we have considered different possible scenarios
predicted by the four site Higgsless model. There, the resonances are not
equally spaced; so we have studied the spectrum, going from very spaced cases
to almost degenerate ones.

The outcome is that all six extra gauge bosons could be detected at
the LHC over a large portion of the parameter space (i.e. couplings
and masses) for a luminosity at regime. The low-edge mass spectrum,
below 1 TeV, could be already discovered during the early stage of
the LHC data taking, with a start-up luminosity of the order of
$L\simeq 1~{\rm fb}^{-1}$.

These results do not include the detector simulation. At present, this work
is in progress, but we do not expect a drastic change in our conclusions
owing to the very clean signals.

\appendix

\section{Gauge boson spectrum}
\label{appendixA}

The charged gauge boson mass Lagrangian is given by:
\begin{equation}\label{LMC}
{\mathcal L}^{{\mathcal C}}_{mass} = \tilde{{\mathcal C}}^-\!
{\mathcal M}^2_c\, \tilde{{\mathcal C}}^+
\end{equation}
with
$
\tilde{{\mathcal C}^-}\!=\left(\!
\begin{array}{ccc}
\Wmt \!\!, &\!\! \Amt\!\!, &\! \!\Bmt\\
\end{array}
\!\right) $ \be \label{M2} {\mathcal M}^2_c=\left(
\begin{array}{ccc}
 \gt^2 f_1^2& - \gt  g_1 f_1^2& 0 \\
-\gt g_1f_1^2 &g_1^2  (f_1^2 +f_2^2)  &  - g_1^2 f_2^2\\
0 & -g_1^2f_2^2 &g_1^2 ( f_1^2 +f_2^2) \\
\end{array}
\right) \ee The eigenvalues for large $g_1$, neglecting ${\mathcal
O}(\gt^4/g_1^4)$, have the following expressions: \be\label{Mw}
M_W^2\approx\tilde{M}_W^2\left(1- \frac{\gt^2}{g_1^2}z_W\right) \ee
\be\label{M1c} M_{1,c}^2\approx M_1^2\left(1 +
\frac{\gt^2}{2g_1^2}\right)
\ee
\be\label{M2c} M_{2,c}^2 \approx
M_2^2\left(1 +\frac{\gt^2}{2g_1^2}z^4\right)
\ee
with the
identifications \be\label{Mwtilde} \tilde{M}_W^2=\gt^2 \frac{f_1^2
f_2^2}{f_1^2 + 2 f_2^2} \quad z_W=\frac{f_1^4 + 2 f_1^2 f_2^2 + 2
f_2^4}{{(f_1^2 + 2 f_2^2)}^2}=\frac{1}{2}(1+z^4) \ee \be\label{A7}
M_1^2=f_1^2 g_1^2 \quad M_2^2=g_1^2(f_1^2 + 2 f_2^2) \quad
z=\frac{M_1}{M_2}=\frac{f_1}{\sqrt{f_1^2+2f_2^2}} \ee We give also,
at the same order, the transformations which express the fields
$\tilde{W}^\pm$, $\tilde{A_1}^\pm$ and $\tilde{A_2}^\pm$
 in terms of the mass eigenvalues
\begin{eqnarray}\label{W}
\tilde{W}^\pm &=&\left(1- \frac{\gt^2}{g_1^2}\frac{z_W}{2}\right) W^\pm \nonumber \\
& &-\frac{\gt}{\sqrt{2} g_1}\left(1 +\frac{\gt^2}{4g_1^2}\frac{1-3z^2}{1-z^2}\right) W_{1}^\pm\nonumber\\
& &-\frac{z^2 \gt}{\sqrt{2} g_1}\left(1+\frac{\gt^2 z^2}{4
g_1^2}\frac{3z^4-5 z^2+4}{1-z^2}\right) W_{2}^\pm
\end{eqnarray}
\begin{eqnarray}\label{A1c}
\tilde{A_1}^\pm &=&\frac{1}{\sqrt{2}}\left(1- \frac{ \gt^2}{4 g_1^2}\frac{1+z^2}{1-z^2}\right) W_{1}^\pm\nonumber \\
& & + \frac{1}{\sqrt{2}}\left(1+\frac{\gt^2 z^4}{4 g_1^2}\frac{1+z^2}{1-z^2}\right) W_{2}^\pm\nonumber \\
& & +\frac{(1+z^2) \gt}{2 g_1} \left(1+\frac{\gt^2 }{4g_1^2}\frac{(1-3z^2)(1+z^4)}{(1+z^2) }\right) W^\pm
\end{eqnarray}
\begin{eqnarray}\label{A2c}
\tilde{A_2}^\pm &=&\frac{1}{\sqrt{2}}\left(1-\frac{\gt^2 }{4 g_1^2}\frac{1-3z^2}{1-z^2}\right) W_{1}^\pm\nonumber \\
& & -\frac{1}{\sqrt{2}}\left(1- \frac{\gt^2 z^4}{4 g_1^2}\frac{3-z^2}{1-z^2}\right) W_{2}^\pm\nonumber \\
& & + \frac{(1-z^2) \gt}{2g_1} \left(1+\frac{\gt^2 }{4 g_1^2} (1-3z^4)\right) W^\pm
\end{eqnarray}
Similarly for the neutral gauge bosons the mass Lagrangian is:
\begin{equation}\label{Lmn}
{\lag}^{\mathcal N}_{mass} = \frac{1}{2}\tilde{N^T} {\mathcal M}^2_n
\tilde{N}
\end{equation}
 with $\tilde{N}^T = \left( \tilde{W^3},
\tilde{Y},
\tilde{A_1^3},
\tilde{A_2^3}\right )$ and
\begin{equation}
\tilde{M}_n^2=\left(
\begin{array}{cccc}
  \gt^2f_1^2& 0&-  \gt g_1 f_1^2 & 0\\
 0 &  (\gt \tan \tilde{\theta})^2f_1^2&0 &- g_1 \gt \tan \tilde{\theta}f_1^2\\
- \gt g_1f_1^2& 0 & g_1^2 ( f_1^2 + f_2^2) &- g_1^2f_2^2 \\
0 & -g_1 \gt \tan \tilde{\theta} f_2^2 &- g_1^2f_2^2 & g_1^2( f_2^2 + f_1^2 ) \\
\end{array}
\right)
\end{equation}
with
\be \tan \tilde{\theta}= \frac{\gpt}{\gt}
\ee
The corresponding
non zero mass eigenvalues, up to ${\mathcal O}(\gt^4/g_1^4)$, are:
\be\label{A14} M_Z^2=\tilde{M}_Z^2\left(1-
\frac{\gt^2}{g_1^2}z_Z\right)
\ee
\be
M_{1,n}^2=M_1^2\left(1
+\frac{\gt^2}{g_1^2} \frac{ \sec^2 \tilde{\theta} }{2}\right) \ee
\be M_{2,n}^2=M_2^2 \left(1+\frac{\gt^2}{g_1^2} \frac{z^4 \sec^2
\tilde{\theta} }{2 }\right) \label{eq:M2}\ee where \be
\tilde{M}_Z^2=\frac{\tilde{M}_W^2}{\cos^2 \tilde{\theta}},
~~~~~z_Z=\frac{1}{2}\frac{(z^4+\cos^22\tilde{\theta})}{\cos^2\tilde{\theta}}
\ee
with the corresponding transformations ($A$ is the photon
field):
\begin{eqnarray}\label{W30}
\tilde{W}^3&=&\cos \tilde{\theta} \left(1-\frac{\gt^2}{4 g_1^2 \cos^2\tilde{\theta}}\left(1+z^4-4 \sin^4\tilde{\theta}\right)\right)  Z \nonumber \\
& & -\frac{\gt}{\sqrt{2}g_1} \left(1-\frac{\gt^2}{4 g_1^2 \cos^2\tilde{\theta}}\left(1-\frac{2(1-2z^2)\cos 2\tilde{\theta}}{1-z^2}\right) \right)Z_{1} \nonumber \\
& & -\frac{\gt z^2}{\sqrt{2}g_1}\left(1+\frac{\gt^2z^2}{4 g_1^2 \cos^2\tilde{\theta}}\left(2-3z^2+\frac{2\cos 2\tilde{\theta}}{1-z^2}\right) \right)
Z_{2} \nonumber \\
& &+ \sin \tilde{\theta}\left(1-\frac{\gt^2 \sin^2\tilde{\theta}}{g_1^2}\right) A
\end{eqnarray}
\begin{eqnarray}\label{Y30}
\tilde{Y} &=&-\sin \tilde{\theta}\left(1-\frac{\gt^2}{4 g_1^2 \cos^2\tilde{\theta}}\left(1+z^4-4\cos^4\tilde{\theta}\right) \right)Z\nonumber \\
& & -\frac{\gt \tan \tilde{\theta}}{\sqrt{2}g_1} \left(1-\frac{\gt^2}{4 g_1^2 \cos^2\tilde{\theta}}\left(1+\frac{2(1-2z^2)\cos 2\tilde{\theta}}
{1-z^2}\right)\right) Z_1\nonumber \\
& &+ \frac{\gt z^2\tan \tilde{\theta}}{\sqrt{2}g_1} \left(1+\frac{\gt^2z^2}{4 g_1^2 \cos^2\tilde{\theta}}\left(2-3z^2-
\frac{2\cos 2\tilde{\theta}}{1-z^2}\right)\right)Z_2 \nonumber \\
& &+ \cos \tilde{\theta}\left(1-\frac{\gt^2 \sin^2\tilde{\theta}}{g_1^2}\right) A
\end{eqnarray}
\begin{eqnarray}\label{A130}
\tilde{A}_{1}^3 &=& \frac{\gt (z^2+\cos 2\tilde{\theta})}{2 g_1\cos \tilde{\theta}} \left(1-\frac{\gt^2}{4 g_1^2 \cos^2\tilde{\theta}}\right.\nn\\
&&\left.\left(\frac{3z^6-\cos^32\tilde{\theta}-(1+2\sin^2\tilde{\theta})z^4-(1-4\cos^4\tilde{\theta})z^2)}{(z^2+\cos 2\tilde{\theta})}
\right)\right)Z \nonumber \\
& &+\frac{1}{\sqrt{2}}\left(1-\frac{\gt^2}{4 g_1^2 \cos^2\tilde{\theta}}\left(1+\frac{2z^2\cos 2\tilde{\theta}}
{1-z^2}\right)\right)Z_{1}\nonumber \\
& &+\frac{1}{\sqrt{2}}\left(1-\frac{\gt^2 z^4}{4 g_1^2 \cos^2\tilde{\theta}}\left(1-\frac{2\cos 2\tilde{\theta}}{1-z^2}\right)\right)Z_{2} \nonumber \\
& &+\frac{\gt\sin \tilde{\theta}}{g_1}\left(1-\frac{\gt^2 \sin^2\tilde{\theta}}{g_1^2}\right) A
\end{eqnarray}
\begin{eqnarray}\label{A230}
\tilde{A}_2^3 &=&-\frac{\gt (z^2-\cos 2\tilde{\theta})}{2 g_1\cos \tilde{\theta}} \left(1-\frac{\gt^2}{4 g_1^2 \cos^2\tilde{\theta}}\right.\nn\\
&&\left.\left(\frac{3z^6-(1+2\cos^2\tilde{\theta})z^4+\cos^32\tilde{\theta}-(1-4\sin^4\tilde{\theta})z^2}{(z^2-\cos 2\tilde{\theta})} \right)\right)Z\nonumber \\
& &+ \frac{1}{\sqrt{2}}\left(1-\frac{\gt^2}{4 g_1^2 \cos^2\tilde{\theta}}\left(1-\frac{2z^2\cos 2\tilde{\theta}}{1-z^2}\right)\right) Z_{1}\nonumber \\
& &-\frac{1}{\sqrt{2}}\left(1-\frac{\gt^2 z^4}{4 g_1^2 \cos^2\tilde{\theta}}\left(1+\frac{2\cos 2\tilde{\theta}}{1-z^2}\right)\right)Z_{2} \nonumber \\
& & +\frac{\gt\sin \tilde{\theta}}{g_1}\left(1-\frac{\gt^2 \sin^2\tilde{\theta}}{g_1^2}\right)A
\end{eqnarray}

\section{Numerical values}
\label{numbers}

As mentioned in \refse{se:boson_production}, we show results for
three values of the $z$-parameter, $z=(0.4,1/ \sqrt{3}, 0.8)$, and
various $b_{1,2}$-sets. For the charged and neutral sectors, the
corresponding mass eigenvalues, widths and fermion-boson couplings are
summarized in the following Tables:
\begin{table}[h]
\begin{center}
\begin{tabular}{|c|c|c|c|c|}
\hline
& $M_{(1,2),c}(\GeV)$& $\Gamma_{1,2} (\GeV)$& $a_1^c$&$a_2^c$\\
\hline \hline
1&508,1251&6.2,35.5&0.03&0.20\\
\hline
2&508,1251&6.2,28.9&-0.01&-0.03\\
\hline
3&1745,3001&183,746&0.15&0.47\\
\hline
4&1745,3001&183,734&-0.15&-0.44\\
\hline
5&1009,1255&35.3,30.5&0.14&0.24\\
\hline
6&1009,1255&33.1,22.2&-0.06&-0.09\\
\hline
\end{tabular}
\end{center}
\caption{Masses, total widths and couplings to SM fermions of the charged
extra gauge bosons for the six scenarios of Table \ref{tab:scenarios}. }
\label{tab1}
\end{table}

\begin{table}[h]
\begin{center}
\begin{tabular}{|c|c|c|c|c|c|c|c|c|c|c|c|c|c|c|}
\hline &$M_{(1,2),n} (\GeV)$&$\Gamma_{1,2} (\GeV)$& ${\hat
a}_{1L}^e$ & ${\hat a}_{1R}^e$ & ${\hat a}_{1L}^d$ & ${\hat
a}_{1R}^d$ & ${\hat a}_{1L}^u$ & ${\hat a}_{1R}^u$ & ${\hat
a}_{2L}^e$ & ${\hat a}_{2R}^e$ & ${\hat a}_{2L}^d$ & ${\hat
a}_{2R}^d$
& ${\hat a}_{2L}^u$ & ${\hat a}_{2R}^u$ \\
\hline \hline
1&510,1251 &6.4,36.0&0.12&0.11&0.05&0.04&-0.09&-0.07&0.43&-0.02&0.46&-0.01&-0.44&0.01\\
\hline
2&510,1251&6.3,28.8&0.02&0.11&-0.05&0.04&0.01&-0.07&-0.08&-0.02&-0.07&-0.01
&0.07&0.01\\
\hline
3&1736,3001&184,756&0.36&0.04&0.34&0.01&-0.35&-0.02&1.06&-0.01&1.07&0.0&-1.07&0.01\\
\hline
4&1736,3001&184,742&-0.32&0.04&-0.34&0.01&0.33&-0.02&-1.0&-0.01&-0.99&0.0
&1.0&0.01\\
\hline
5&1012,1256 &36.2,32.0&0.37&0.08&0.31&0.03&-0.34&-0.06&0.50&-0.05&0.54&-0.02&-0.52&0.04\\
\hline
6&1012,1256&33.7,22.9&-0.11&0.08&-0.16&0.03&0.14&-0.06&-0.24&-0.05&-0.20&-0.02&0.22&0.04\\
\hline
\end{tabular}
\end{center}
\caption{Masses, total widths and couplings to SM fermions
of the neutral extra gauge bosons for the six scenarios of Table
\ref{tab:scenarios}. The electric charge (-e) is factorized.}
\label{tab4}
\end{table}
{\bf Acknowledgements:}
 E. Accomando  would like to thank the
Department of Physics of the University of Florence for partial
financial support. We thank A. De Roeck for valuable discussions and
suggestions, E. Migliore for useful informations and R. Contino for comments
on LEP2 limits. D. Dominici was partially supported by MIUR under the contract PRIN-2006020509.

\bibliography{foursites}

\end{document}